%% file: main.tex
\documentclass[acmsmall,screen]{acmart}

\AtBeginDocument{%
  \providecommand\BibTeX{{%
    \normalfont B\kern-0.5em{\scshape i\kern-0.25em b}\kern-0.8em\TeX}}}

\setcopyright{acmcopyright}
\copyrightyear{2021}
\acmYear{2021}
\acmDOI{10.1145/1122445.1122456}

\acmJournal{JACM}

\usepackage[ruled,vlined,lined,commentsnumbered]{algorithm2e}
\usepackage{booktabs}
\usepackage{moreverb}
\usepackage{fontenc}
\usepackage{amsmath}
\usepackage{fancybox}
\usepackage{wrapfig}
\usepackage{color}
\usepackage{colortbl}
\usepackage{array}
\usepackage{multirow}
\usepackage{multicol}
\usepackage{listings}
\usepackage{makecell}
\usepackage{graphicx}
\usepackage{rotating}
\usepackage{setspace}
\usepackage{soul}
\usepackage{balance}
\usepackage{csvsimple}
\usepackage{longtable}
\usepackage[skip=0pt,labelfont=bf]{caption}
\usepackage{url}
\usepackage{lscape}
\usepackage{rotating}
\usepackage{tikz}
\usepackage{enumitem}
\usepackage{subcaption}

\usepackage[most]{tcolorbox}

\usepackage{threeparttable}

\usepackage{bclogo}

\usepackage{marvosym}
\usepackage{pifont}

\newcolumntype{L}[1]{>{\raggedright\let\newline\\\arraybackslash\hspace{0pt}}m{#1}}
\newcolumntype{C}[1]{>{\centering\let\newline\\\arraybackslash\hspace{0pt}}m{#1}}
\newcolumntype{R}[1]{>{\raggedleft\let\newline\\\arraybackslash\hspace{0pt}}m{#1}}

\definecolor{codegreen}{rgb}{0,0.6,0}
\definecolor{codegray}{rgb}{0.5,0.5,0.5}
\definecolor{codepurple}{rgb}{0.58,0,0.82}
\definecolor{backcolour}{rgb}{0.95,0.95,0.92}

\lstdefinestyle{mystyle}{
    commentstyle=\color{codegreen},
    keywordstyle=\color{magenta},
    numberstyle=\tiny\color{codegray},
    stringstyle=\color{codepurple},
    basicstyle=\footnotesize,
    breakatwhitespace=false,
    breaklines=true,
    captionpos=b,
    keepspaces=true,
    showspaces=false,
    showstringspaces=false,
    showtabs=false,
    tabsize=2
}

\setlist{noitemsep} 

\definecolor{darkpastelred}{rgb}{0.76, 0.23, 0.13}
\definecolor{ao(english)}{rgb}{0.0, 0.5, 0.0}

\definecolor{darkpastelred}{rgb}{0.76, 0.23, 0.13}
\definecolor{ao(english)}{rgb}{0.0, 0.5, 0.0}

\lstdefinelanguage{diff}{
  morecomment=[f][\color{blue}]{@@},     
  morecomment=[f][\color{red}]-,         
  morecomment=[f][\color{codegreen}]+,       
  morecomment=[f][\color{red}]{---}, 
  morecomment=[f][\color{codegreen}]{+++},
}

\definecolor{yellow}{RGB}{255,255,153}
\definecolor{grey}{RGB}{224,224,224}

\newboolean{showcomments}
\setboolean{showcomments}{true}
\ifthenelse{\boolean{showcomments}}
 { \newcommand{\mynote}[2]{
      \fbox{\bfseries\sffamily\scriptsize#1}
        {\small$\blacktriangleright$\textsf{\emph{#2}}$\blacktriangleleft$}}}
        { \newcommand{\mynote}[2]{}}

\definecolor{DarkOrange}{rgb}{0.8,0.3,0.0}
\definecolor{DarkCyan}{rgb}{0.0, 0.55, 0.55}

\newcommand{\etal}{\emph{et~al.}\xspace}

\newcolumntype{?}{!{\vrule width 1pt}}

\newcommand*{\eg}{e.g., }
\newcommand*{\ie}{i.e., }

\newcommand*{\mycode}{\fontfamily{lmtt}\selectfont}

\newcommand{\toolname}{\textsc{Beep}\xspace}
\newcommand{\repairtoolname}{\texttt{PEARL}\xspace}
\newcommand{\ACG}{\texttt{AnyCodeGen}\xspace}

\newcommand{\update}[1]{\textcolor{blue}{#1}}

\newcommand{\intuition}[1]{
\begin{tcolorbox}[tile,size=fbox,boxsep=1mm,boxrule=0pt,top=0pt,bottom=0pt,
borderline west={1mm}{-2pt}{black!50!white},colback=black!5!white]
\em #1
\end{tcolorbox}
}

\newcommand{\notez}[1]{
\begin{tcolorbox}[tile,size=fbox,boxsep=1.1mm,boxrule=0pt,top=0pt,bottom=0pt,colback=black!5!white]
\em #1
\end{tcolorbox}
}

\begin{document}

\title{\toolname: Fine-grained Fix Localization by Learning to Predict
             Buggy Code Elements}

\author{Shangwen Wang}
\email{wangshangwen13@nudt.edu.cn}
\affiliation{%
   \institution{National University of Defense Technology}
 	\city{Changsha}
 	\country{China}
}

\author{Kui Liu}
\authornote{Corresponding author.}
\email{kui.liu@nuaa.edu.cn}
\affiliation{%
   \institution{Nanjing University of Aeronautics and Astronautics}
 	\city{Nanjing}
 	\country{China}
}

\author{Bo Lin}
\email{linbo19@nudt.edu.cn}
\affiliation{%
   \institution{National University of Defense Technology}
 	\city{Changsha}
 	\country{China}
}

\author{Li Li}
\email{li.li@monash.edu}
\affiliation{%
  \institution{Monash University}
  \city{Melbourne}
  \country{Australia}
}

\author{Jacques Klein}
\email{jacques.klein@uni.lu}
\affiliation{%
  \institution{University of Luxembourg}
  \country{Luxembourg}
}

\author{Xiaoguang Mao}
\email{xgmao@nudt.edu.cn}
\affiliation{%
   \institution{National University of Defense Technology}
 	\city{Changsha}
 	\country{China}
}

\author{Tegawend\'e F. Bissyand\'e}
\email{tegawende.bissyande@uni.lu}
\affiliation{%
  \institution{University of Luxembourg}
  \country{Luxembourg}
}

\input{0.abstract}



\maketitle

\input{1.intro}
\input{2.background}

\input{4.model}

\input{5.setup}

\input{6.evaluation}

\input{7.discussion}
\input{8.relatedwork}
\input{9.conclusion}


\bibliographystyle{ACM-Reference-Format}
\bibliography{bib/references}

\end{document}

%% file: 0.abstract.tex
\begin{abstract} 

Software Fault Localization refers to the activity of finding code elements (e.g., statements) that are related to a software failure. The state-of-the-art fault localization techniques, however, produce coarse-grained results that can deter manual debugging or mislead automated repair tools. In this work, we focus specifically on the fine-grained identification of code elements (i.e., tokens) that must be changed to fix a buggy program: we refer to it as {\em fix localization}. This paper introduces a neural network architecture (named \toolname) that builds on AST paths to predict the buggy code element as well as the change action that must be applied to repair a program. Leveraging massive data of bugs and patches within the CoCoNut dataset, we trained a model that was (1) effective in localizing the buggy tokens with the Mean First Rank significantly higher than a statistics based baseline and a machine learning-based baseline, and (2) effective in predicting the repair operators (with the associated buggy code elements) with a Recall@1$\approx$ 30-45\% and the Mean First Rank $\approx$7-12 (evaluated by CoCoNut, ManySStuBs4J, and Defects4J datasets). 
To showcase how fine-grained fix localization can help program repair, we employ it in two repair pipelines where we use either a code completion engine to predict the correct token or a set of heuristics to search for the suitable donor code.
A key strength of accurate fix localization for program repair is that it reduces the chance of patch overfitting, a challenge in generate-and-validate automated program repair: both two repair pipelines achieve a correctness ratio of 100\%, i.e., all generated patches are found to be correct. 
Moreover, accurate fix localization helps enhance the efficiency of program repair.
\end{abstract}

%% file: 1.intro.tex
\section{Introduction}
\label{sec:intro}

The complexity of modern software is a source of bugs (leading to program failures), which makes debugging a resource-intensive activity of software development. Automating debugging has thus become a core research field of computer science. Mainly, debugging activities are twofold: fault localization and software repair \cite{lou2020can}. While there is currently excitement in the research around the automation of software repair, fault localization has always been regarded as one of the most tedious and time-consuming activities in debugging~\cite{jones2002visualization}. 
There is a large amount of literature~\cite{wong2016survey} on advanced fault localization techniques that are developed to aid software engineers to locate program bugs. The granularity and accuracy of such techniques remain however a key bottleneck for their adoption by developers in the industry~\cite{parnin2011automated}. Research has indeed shown that literature approaches to fault localization are limited by a set of strong assumptions on how developers behave when debugging \cite{parnin2011automated}. The researchers revealed that examining a faulty statement is not enough for a developer to understand and fix the corresponding bug. We foresee two different approaches to cope with this limitation: (1) provide a precise characterization of bug alongside coarse-grained localization information (e.g., buffer overflow bug in line~$x$); or (2) provide fine-grained localization of code elements to change alongside the required change operator (e.g., UPDATE boolean value ``true" in line~$x$, column~$y$). Our work focuses on the latter.

With the momentum of automated program repair (APR), in particular with the promising generate-and-validate approaches, fault localization techniques are widely leveraged to automate the APR pipeline in terms of identifying code locations that must be involved with code changes to generate patches. Unfortunately, on the one hand, the granularity (e.g., method or line) of fault localization outputs is such that APR tools contribute to exploding the search space in patch generation, leading to an inefficient repair process~\cite{liu2020efficiency}. On the other hand, the inaccuracy of fault localization tools leads APR tools to generate plausible patches (\ie patches that pass the available test suites without necessarily being correct) that are applied to the non-buggy locations. 
This is commonly known as the overfitting problem~\cite{wang2020automated,tian2020evaluating,xiong2018identifying,smith2015cure,long2016analysis} (\ie a plausible patch may still be incorrect), which is now regarded as a key bottleneck for APR adoption in the industry \cite{tao2014automatically}. Finally, a recent study~\cite{liu2019you} has highlighted that the poor performance of state-of-the-art APR tools on benchmark defects can be largely attributed to the under-performance of fault localization: about 1 out of 3 real-world bugs in the Defects4J benchmark~\cite{just2014defects4j} cannot be localized accurately by commonly-used spectrum-based fault localization~\cite{liu2019you}. Overall, fault localization is a critical concern towards facilitating both manual and automatic bug fixing.

Consider the example of Defects4J bug Closure-62 as well as its patch shown in Figure~\ref{fig:closure-62}. Although a fault localization tool may localize the buggy {\mycode if} statement, developers still need to investigate a large number of change trials for addressing the bug. Similarly, a typical search-based generate-and-validate APR tool needs to make various change trials on 12 single code elements (i.e., 12 code tokens: ``{\tt excerpt}'', ``{\tt equals}'', ``{\tt LINE}'', ``{\tt \&\&}'', ``{\tt 0}'', ``{\tt <=}'', ``{\tt charno}'', ``{\tt \&\&}'', ``{\tt charno}'', ``{\tt <}'', ``{\tt sourceExcerpt}'' and ``{\tt length}'').
On the one hand, conducting trials on non-buggy code elements will generate and validate a number of nonsensical patches, hence impacting efficiency due to expensive test campaigns~\cite{liu2020efficiency}.
On the other hand, modifying the non-buggy code elements increases the likelihood of yielding overfitting patches. For this bug,  jKali~\cite{martinez2016astor} removes the whole conditional expression as the faulty element, which is sufficient to pass the weak test suite.

\begin{figure}[!t]
	\centering
\begin{lstlisting}[language=diff, firstnumber=1, numbers = left,xleftmargin=2em, frame=lines,]
// Ground-truth patch for Closure-62
-    if (excerpt.equals(LINE) && 0 <= charno && charno < sourceExcerpt.length()) {
+    if (excerpt.equals(LINE) && 0 <= charno && charno <= sourceExcerpt.length()) {

// An overfitting patch generated for Closure-62 by jKali
-    if (excerpt.equals(LINE) && 0 <= charno && charno < sourceExcerpt.length()) {
+    if (true) {
\end{lstlisting}
	\caption{The ground-truth patch and an overfitting patch from jKali for the bug Closure-62.}
  \label{fig:closure-62}
\end{figure}

Our work aims to improve fault localization for debugging, with a focus on refining the granularity of code elements to localize, without sacrificing accuracy. We start with the common assumption that underlies most machine  learning techniques to defect prediction~\cite{li2019improving}: buggy programs having similar program structures tend to involve the same buggy elements. We, therefore, propose to leverage the concept of abstract syntax tree (AST) path \cite{alon2018general}, which was demonstrated amenable for learning program structure information
~\cite{alon2019code2vec,alon2019code2seq,alon2020structural}. Each code element can thus be accurately identified in the program structure through its AST path.
Eventually, we propose a learning approach for fine-grained identification of which code elements must be changed in a buggy program: we refer to this as {\bf fix localization}.
Our ambition with fix localization, therefore, goes beyond current fault localization approaches circumscribing the code locations (i.e., at best, the statements) suspected of causing a failure.
We develop a new fix localization approach, \toolname (\underline{B}uggy codE \underline{E}lements \underline{P}redictor), which takes as inputs a faulty method and yields a ranked list of code elements that are likely to be the buggy ones. 

We performed extensive experiments to assess the performance of \toolname. Specifically, we trained \toolname on the dataset used by CoCoNut \cite{lutellier2020coconut} repair tool, and evaluated the prediction results on three defect benchmarks in total: CoCoNut (by using 10-fold cross validation), ManySStuBs4J \cite{karampatsis2020how}, and Defects4J \cite{just2014defects4j}.
Since previous studies~\cite{liu2019you} have highlighted that spectrum-based fault localization yields reasonable performance at coarse granularity (e.g., file or method), we consider a fine-grained localization scenario where the buggy method is known. We also perform experiments in a scenario where the faulty line is provided (e.g., by another tool).
Overall, the experimental results reveal that the buggy code elements identified by \toolname can significantly reduce the debugging effort. 
For instance, the median token number for the buggy method in the ManySStuBs4J dataset is 72 while \toolname can predict the buggy token in top-7 on average. 
Moreover, \toolname is also capable of accurately predicting the code change operator associated to the buggy element.

To assess the performance on fixing bugs with the predicted buggy code elements, we design two repair pipelines: in the first (1) we leverage an off-the-shelf code completion tool to predict the correct contents for tokens identified as buggy by \toolname; and in the second (2) we use straightforward heuristics to search for replacements for the localized buggy tokens. Experimental evaluation of the repair pipelines on existing defect benchmarks (Bears, Defects4J, Bugs.jar, and QuixBugs) reveals that these pipelines can fix 27 and 32 bugs respectively, both with a 100\% correctness ratio. 
Moreover, the efficiency of the pipeline (measured by the Number of Patch Candidates, NPC) significantly outperforms those of the state-of-the-art techniques in the literature (\eg SimFix~\cite{jiang2018shaping} and TBar~\cite{liu2019tbar}).

To sum up, this paper makes the following contributions:
\begin{itemize}[leftmargin=*]

    \item[\ding{172}] [{\bf Fine-grained fix localization with \toolname}]: We develop a new fine-grained fix localization approach, \toolname, 
    which takes advantage of a specialized deep learning architecture that considers abstract syntax tree paths to precisely predict buggy AST nodes. 
    Experimental results with extensive patch datasets (ManySStuBs4J, CoCoNut, and Defects4J) demonstrate the effectiveness of our prediction model which precisely predicts at Top-1 the buggy code elements for up to 45\% bugs. An ablation study further demonstrates the importance of the various components and design decisions in our model.   
    
    \item[\ding{173}] [{\bf Prediction of code change operators for identified buggy code elements}]: In the implementation of the \toolname fix localization approach, we also learn to predict which operators (i.e., {\tt DELETE}, {\tt UPDATE} and {\tt INSERT}) should be applied to the localized buggy code elements, in order to fix the program. Evaluation results reveal that \toolname can also accurately predict the associated code change operator.
    \item [\ding{174}] [{\bf Demonstration of \toolname added-value in repair processes}]: 
    Given the output of \toolname (i.e., an identified buggy code element and its associated predicted change operator), we design two repair schemes that leverage either an off-the-shelf code completion tool or a set of straightforward heuristics. We then assess the repair performance of the implemented repair pipelines on four widely-used repair benchmarks. Results reveal that with fine-grained accurate localization, the patch generation performance is characterized by high precision (100\% generated valid patches are also correct) and high efficiency (with an average of only 2 patch candidates to test before finding a valid patch). The pipeline is also complementary to the state-of-the-art since it fixes new bugs that were not yet fixed in the literature. 
\end{itemize}

%% file: 2.background.tex
\section{Background and Definitions}
\label{sec:bg}

\subsection{Automated Program Repair}
APR is a software engineering research field that seeks to automate patch generation towards releasing developers from the heavy burden of manual debugging.
APR pipelines start with a fault localization step that enumerates a list of code locations that are suspected as buggy. In widespread generate-and-validate APR tools, spectrum-based fault localization techniques~\cite{pearson2017evaluating} are adopted to produce bug positions at the line~\cite{liu2020efficiency} or method~\cite{wen2018context} granularity. 
Because the performance of fault localization can severely impact repair performance~\cite{liu2019you}, researchers act on this step to improve the pipeline: e.g., ACS~\cite{xiong2017precise} uses predicate switching~\cite{zhang2006locating} while SimFix~\cite{jiang2018shaping} applies a test case purification approach~\cite{xuan2014test} to refine the fault localization results, respectively. In recent works~\cite{lutellier2020coconut,jiang2021cure}, the assessment of the actual patch generation step of APR has been done by assuming that the fault localization at the line is perfect. Nevertheless, even in such cases, the patch generation is challenged since it often partially touches the relevant code elements~\cite{liu2018closer}. In summary, state-of-the-art APR approaches are still impacted by the presence of non-buggy code elements within the suspected buggy code lines. Our work explores fix localization to achieve finer-grained localization of buggy code elements for program repair.


\subsection{Automated Fault localization}
Automated fault localization \cite{wong2016survey} aims to precisely diagnose bug positions to facilitate program debugging. 
The most widely studied {\em spectrum-based fault localization} (SBFL) techniques~\cite{pearson2017evaluating} usually utilize the execution traces of passing and failing tests to identify the suspicious code elements (\eg lines/methods). The intuition behind this is that if a code element is covered by more failing tests but fewer passing tests, then it is more likely to be the buggy one.
Hence, researchers usually apply statistical analysis (\eg Ochiai~\cite{abreu2007accuracy} and Tarantula~\cite{jones2002visualization}) to calculate the suspiciousness of code elements and rank them in descending order to represent the exposed bug positions.
The inherent limitation of SBFL approaches is that a code element executed by a failing test does not necessarily indicate that the element is related to the test failure. To solve this problem, researchers also propose {\em mutation-based fault localization} (MBFL) techniques that mutate code elements and then check their impact on the test outcomes~\cite{papadakis2015metallaxis}.
Besides SBFL and MBFL, researchers have explored various fault localization techniques, e.g., slice-based~\cite{zhang2007study}, statistics based~\cite{liblit2005scalable}, program state-based~\cite{zeller2002simplifying}, learning-based~\cite{wong2011effective}, data mining based~\cite{koyuncu2019d}, and model-based techniques~\cite{mayer2008evaluating}.
Nevertheless, the state-of-the-art fault localization techniques provide the identified bug positions at best at the granularity of code lines. Our work is the first to target the identification of buggy code elements nested within buggy lines, thus providing a more fine-grained localization for program debugging.

\subsection{AST Paths for Code Representation}
\label{sec:ast-definition}

\begin{figure}[!ht]
    \centering
    \vspace{1mm}
    \includegraphics[width=0.6\linewidth]{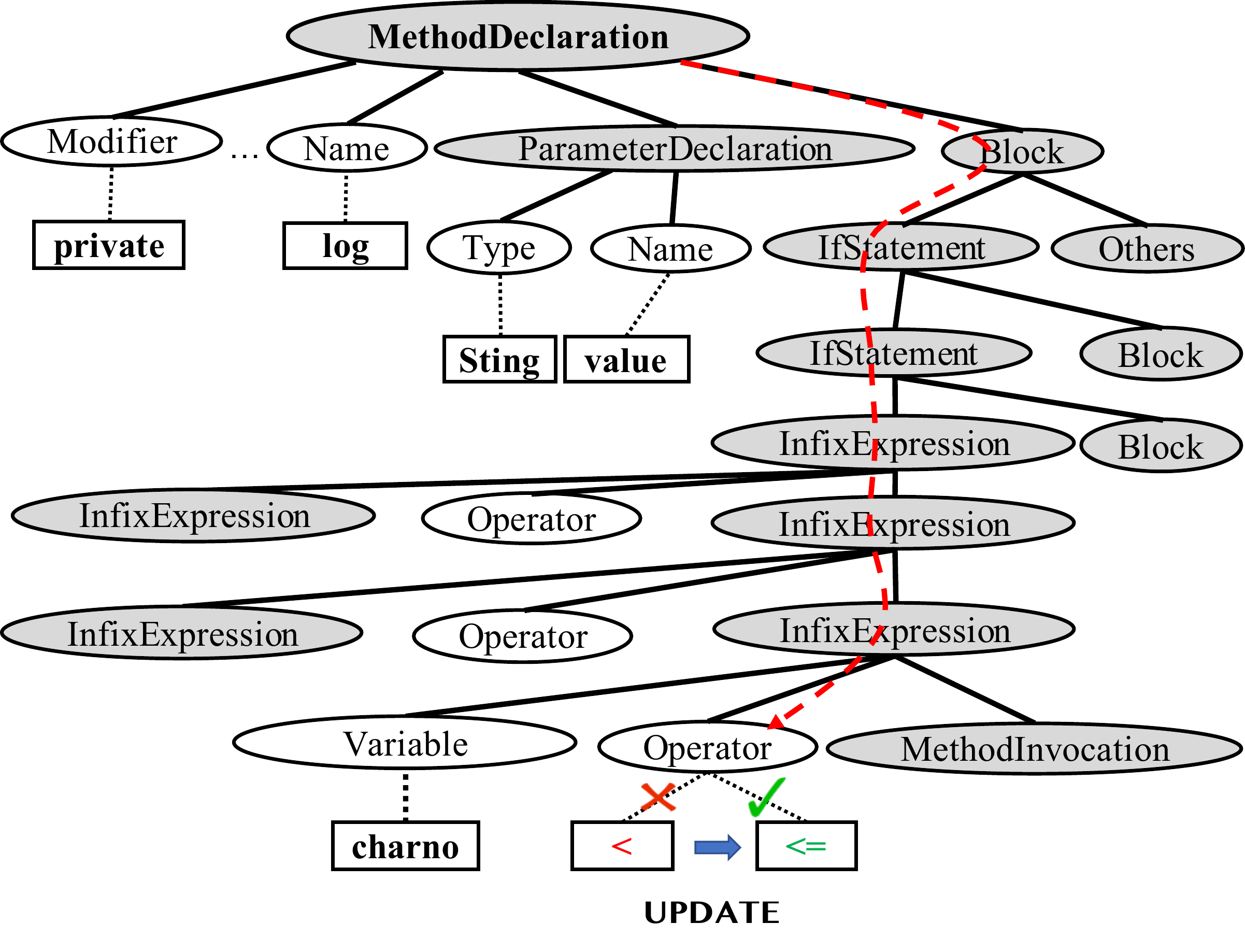}
    \caption{AST \& change operator for the correct patch in Fig.~\ref{fig:closure-62}.}
    \label{fig:ast-path}
\end{figure}

Recently, a number of works consider the Abstract Syntax Tree (AST) path \cite{alon2018general} for code representation. Code2vec~\cite{alon2019code2vec} and Code2seq~\cite{alon2019code2seq} use all the AST paths within a method with an attention mechanism to represent this method and predict its name as well as generate its natural language description.
Brody \etal \cite{brody2020structural} use this technique to encode the changed code within a file and then predict the following code changes.
Alon \etal \cite{alon2020structural} adopt it to address the {\em any-code completion} task, which is generating a missing piece of source code in a given program.
Compton \etal \cite{compton2020embedding} extend the application scenario of this technique to the whole Java class via variable obfuscation.
All the above works demonstrate that the AST path technique is amenable for learning program structure information.
We thus build on the concepts of {\em AST path} and {\em operation path} associated to the buggy code elements (see example in Fig.~\ref{fig:ast-path}). Formal definitions for each term in our study are provided as below.

\noindent
\textbf{Definition 1 - [AST] :} The AST of a code snippet is defined as a tuple:
 $\langle N, L, T, r, \Delta, \Phi\rangle$,
 where $N$ is a set of non-leaf nodes, $L$ is a set of leaf nodes, and $T$ is a set of code tokens for $L$. $r \in N$ represents the root node, $\delta\in\Delta : n\rightarrow n^\prime, n \in N, n^\prime \in (N \cup T)$ is a function that reflects the {\em parent-child} relationship between two AST nodes $n$ and $n^\prime$.
    $\phi\in\Phi : l \rightarrow t, l \in L, t \in T$ maps a leaf node with a corresponding code token. 


\noindent
\textbf{Definition 2 - [AST path] :} An AST path is a path starting from the root node $r$ to a leaf node $l$, that is defined as a quadruple: $p = \langle r, l, N^\prime, \Delta^\prime\rangle, l \in L, N^\prime\subset N, \Delta^\prime\subset\Delta$.


\noindent
\textbf{Definition 3 - [Operation Path] :} An operation path is an AST path associated with code change operator that works on a leaf node $l$, which is defined as a triple: $op = \langle t, p, o\rangle$,
where $t$ is the code token of the leaf node $l$ in the AST path $p$, and $o \in \{\texttt{UPDATE}, \texttt{DELETE}, \texttt{INSERT}\}$ is an atomic code change operator that works on the leaf node. 

Fig.~\ref{fig:ast-path} provides an illustration of the AST path (with the change operator) for the ground-truth patch illustrated in Fig~\ref{fig:closure-62}.
{\tt MethodDeclaration} is the root node of the AST, and other AST nodes with \colorbox{grey}{grey} backgrounds are non-leaf nodes, while all leaf nodes are presented with transparent backgrounds (oval shapes).
Each leaf node is attached with its associated code token (in rectangles).
The AST path for the buggy operator ``{\tt <}'' is $p$: ``{\tt MethodDeclaration $\rightarrow$ Block $\rightarrow$ IfStatement $\rightarrow$ IfStatement $\rightarrow$ InfixExpression $\rightarrow$ InfixExpression $\rightarrow$ InfixExpression $\rightarrow$ Operator }'',
highlighted with a red arrow in Fig.~\ref{fig:ast-path}. For simplification, the other AST nodes irrelevant to the buggy code element are not presented in this figure.
The operation path for fixing the buggy operator is $op = \langle$`$<$'$, p, \texttt{UPDATE}\rangle$.
As shown in Figure~\ref{fig:ast-path}, the buggy operator `$<$' is replaced (i.e., \texttt{UPDATE}) with the operator `$<=$'.

\subsection{Declarations}
\label{sec:declaration}
In this section, we provide the declarations of code elements and fix localization to clarify their differences between this work and existing studies explored in the literature.

\noindent
\textbf{[Code Element]:} Generally speaking, all code entities (such as code fragments, statements, expressions, and single code tokens) in the programs can be referred to as code elements for concrete targets. In this work, \toolname is to perform the fine-grained fault localization at code token level for program repair, thus ``{\it code elements}'' studied in this paper denote code tokens in the program.

\noindent
\textbf{[Fix Localization]:} 
In the basic pipeline of APR, fault localization aims to pinpoint the statement(s) that can be selected as bug locations for mutation~\cite{le2013current}.
The accuracy of fault localization can directly impact the bug-fixing performance of APR tools~\cite{liu2019you}.
In the literature, researchers have taken different efforts to improve the fault localization for automated program repair. 
For example, Tan et al.~\cite{tan2016anti} utilized the anti-patterns to improve fault localization by pinpointing the buggy location at line/function level with higher accuracy. 
Shariffdeen et al.~\cite{shariffdeen2021automated} leveraged the transformation of existing patches to identify the patch insertion points of bugs at line level.
Xin and Reiss~\cite{xin2017leveraging} utilized the stack trace of crashed programs after executing test cases to enhance the fault localization for program repair.
Nevertheless, as stated by Le Goues et al.~\cite{le2013current}, the challenge of fault localization for program repair is ``{\it statements that are executed exclusively by the bug-inducing input may not be the best locations for a repair}.''
Our work explores to address this challenge by predicting the fine-grained fault localization (referred to as fix localization in this work) at code element (i.e., code token that need to be fixed within the buggy program) level for program repair, which is different from the existing efforts achieved in the literature on improving fault localization accuracy at line/function level.

%% file: 4.model.tex
\section{Proposed Approach}
\label{sec:approach}

\begin{figure}[!ht]
	\centering
	\includegraphics[width=0.7\linewidth]{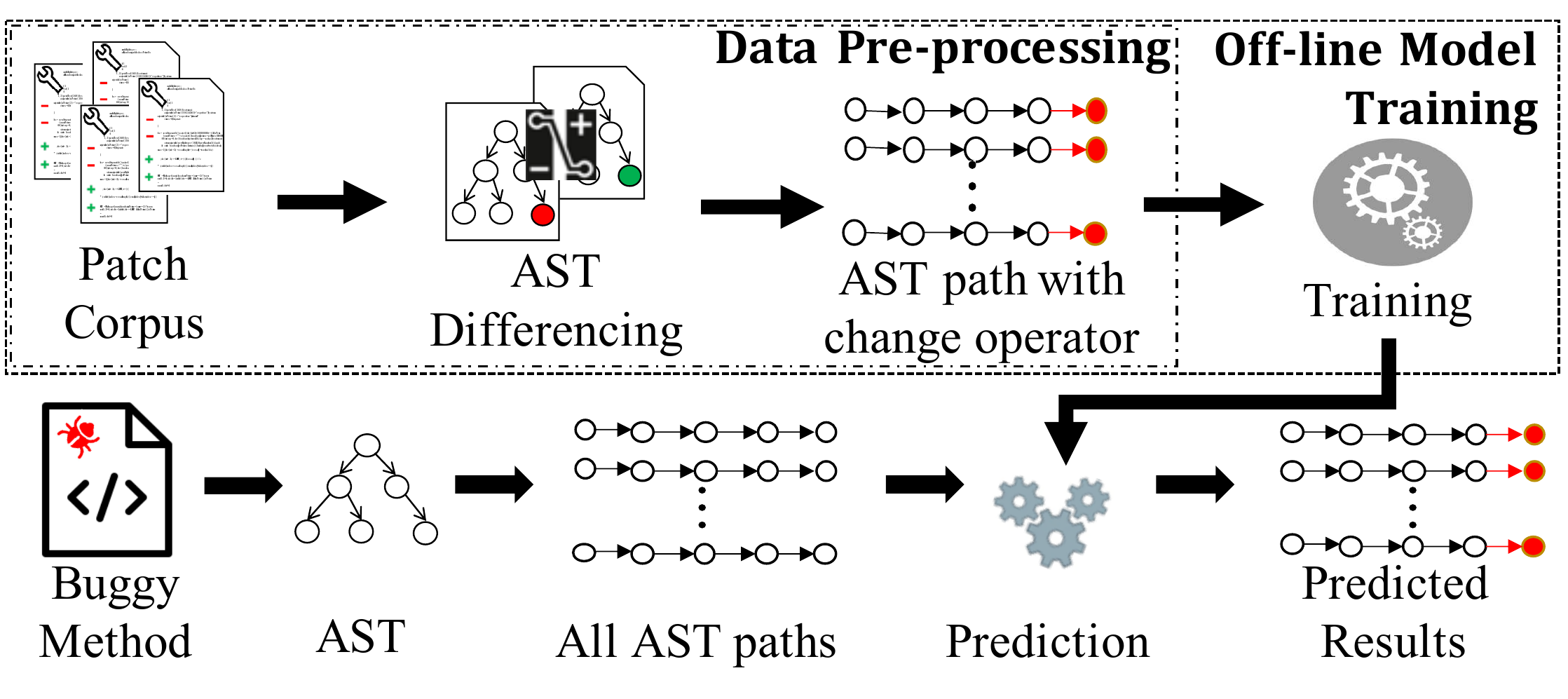}
	\caption{Overview of our proposed \toolname.}
	\label{fig:model}
\end{figure}

\toolname is built based on a neural network model trained offline with patches collected in the wild. The model is then used to predict the buggy code element and the associated code change operator. As illustrated in Figure~\ref{fig:model}, the first step in \toolname is a pre-processing step that produces AST differences as representations of patches.

\subsection{\bf Data pre-processing}
We consider a collection of historical patches as a training dataset. Then, for each, we use the GumTree~\cite{falleri2014fine}  AST differencing algorithm to compute AST diffs, which are the actual input representations for \toolname. These diffs thus allow to readily identify the AST path (from the root towards the modified leaf node) as well as the operation path (i.e., code token + AST path + change operator).
This set, which so far includes only true positive samples, is augmented by considering, for every buggy method, all AST paths towards non-buggy leaf nodes and assigning alternatively each of the three change operators. 
These new operation paths are therefore tagged as true negatives for the training since they represent paths that should not be predicted as buggy.

Additionally, during pre-processing, we collect the code token of the leaf node in each AST path to be included in the feature set for learning.
Following the insights of a recent study by Lutellier~\etal~\cite{lutellier2020coconut}, we split code tokens into sub-token sequences in order to significantly reduce the size of the vocabulary. Concretely, as per previous studies~\cite{alon2019code2seq,allamanis2015suggesting}, code tokens are broken into sub-token sequences ``\{$t_0, t_i,\ldots,t_n$\}'' based on camel case and underscore naming conventions, and the obtained sub-tokens are converted to their lowercase form.



\subsection{\bf Training the \toolname model}
\label{sec:training}
\begin{figure}[!ht]
	\centering
	\includegraphics[width=0.9\linewidth]{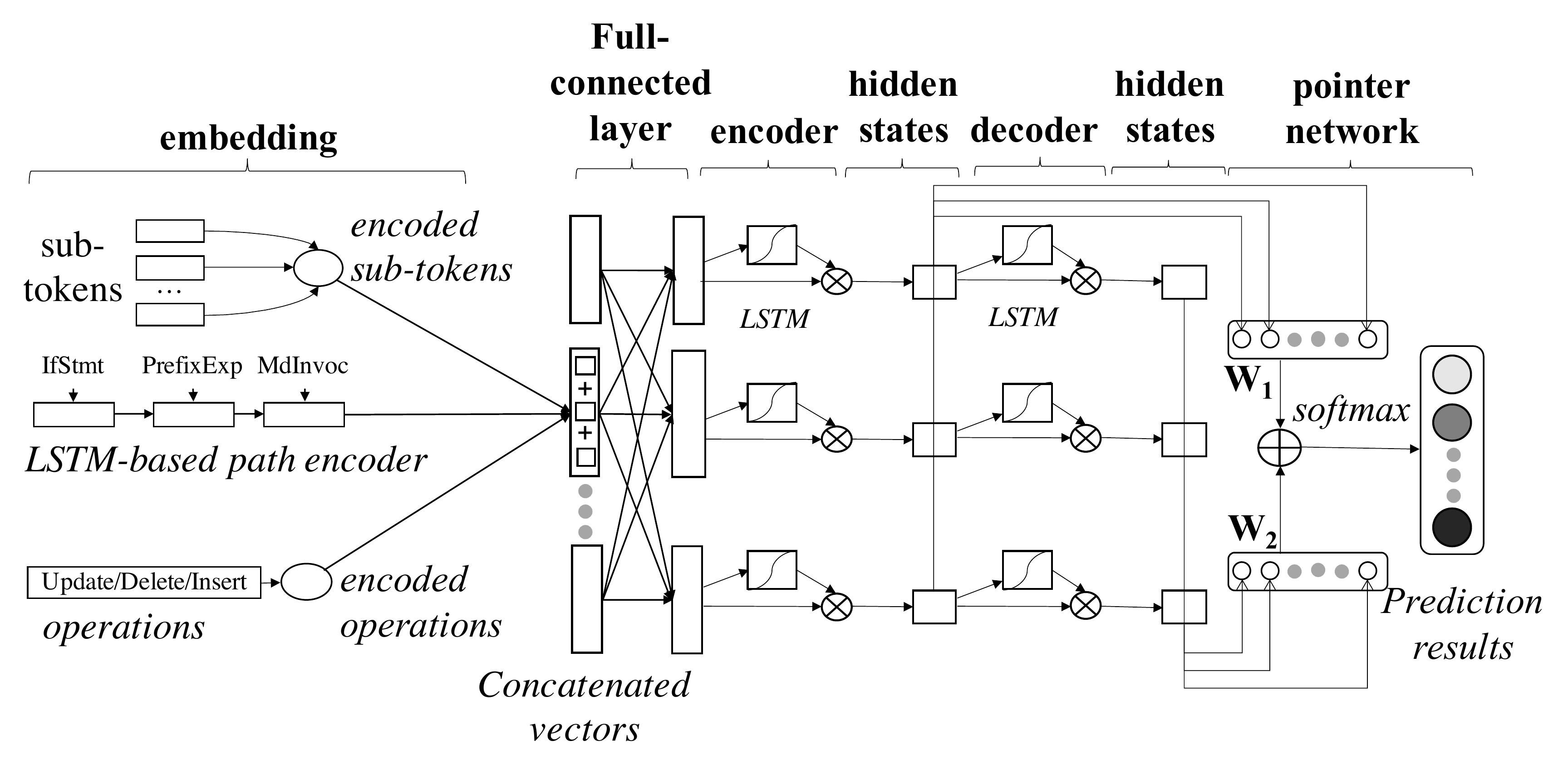}
	\caption{Architecture of the Prediction Model.}
	\label{fig:model2}
\end{figure}

\toolname is mainly composed of an {\em encoder-decoder} network and a {\em pointer network} to generate the output. Our prediction approach is based on the intuition of software naturalness~\cite{hindle2012naturalness}: the buggy part should be detectable via learning as encoder-decoders have already done for natural language typographical/grammatical mistakes.
A pointer network \cite{vinyals2015pointer} is a simple modification of the {\em attention} model which learns the conditional probability of an output sequence with elements that are discrete tokens corresponding to positions in an input sequence. As several recent studies~\cite{brody2020structural,li2020survey} have further demonstrated, pointer networks are indeed particularly effective when the output is simply picked among elements from inputs.
Since in our problem case, the predicted results (i.e., buggy code elements and change operators) are discrete elements, \toolname leverages the pointer network to predict the buggy code elements along with the associated change operators.
Figure~\ref{fig:model2} illustrates the overview of the model architecture.

Given an operation path, \toolname first respectively encodes code sub-tokens, AST path, and operator with three models\footnote{Encoding each of the sub-tokens, AST path, and operations is followed with the encoding method of code2seq~\cite{alon2018code2seq}.}, and concatenates the vectors for path representation. Then, the path vectors are passed through a fully-connected layer and the Long Short-Term Memory (LSTM) based encoder-decoder consecutively. Finally, the pointer network learns to predict the most probable path.


\paragraph{Embedding Operation Paths}
Given a set of $k$ operation paths \{$op_1$,\ldots,$op_k$\}, \toolname embeds a vector representation $v_i$ for each path $op_i$ = $\langle$ $t_i$, $p_i$, $o_i$ $\rangle$ where $p_i$ = \{$n_1^i,n_2^i$,\ldots,$n_{l_i}^i$\} is the corresponding AST path, $t_i$ is the code token and $o_i$ is the change operator. 
To that end, \toolname first leverages a learned embedding matrix to embed each sub-token (after splitting $t_i$) and sum the sub-token vectors to represent the full token. The change operator of each operation path is embedded with another embedding matrix. The embedding process is formulated as below:
\begin{equation}
	\label{eqa:encode-token}
	\begin{aligned}
    & \mbox{V}_{t} = \sum_{t_s \in T_{s}} E_{t}(t_s) \\
    & \mbox{V}_{o} = E_{o}(o) 
    \end{aligned}
\end{equation}
where $E_{t}(*)$ and $E_{o}(*)$ are learned embedding matrix. $T_{s}$ is the sub-token sequence of code token $t$ and $V_t$ represents the vector representation for token $t$,
while $V_o$ is the vector representation for the change operator $o$.


The AST path of each operation path is composed of several AST nodes. \toolname also represents each node $n_i$ using a learned embedding matrix $E_{p}(*)$ and encode the entire sequence with the final state of the bi-directional LSTM neural networks:
\begin{equation}
    \mbox{V}_{p} = LSTM(E_{p}(n_1),E_{p}(n_2),\ldots,E_{p}(n_l))
    \label{eqa:encode-path}
\end{equation}
where $V_p$ represents the vector representation of an AST path. $LSTM$ denotes the bi-directional LSTM neural networks.



\paragraph{Extracting Features with Encoder-Decoder Networks}
To represent the operation path, we jointly concatenate the vector representations of code token, the AST path, and the change operator, and then pass them through a fully-connected layer, of which results are further fed into the LSTM-based encoder-decoder networks to better capture the features, which is formulated as below:
\begin{equation}
    \begin{aligned}
	& z_i = tanh(W_{in}[V_t; V_p; V_o]) \\
    & (e_1,\ldots,e_k) = LSTM_{encoder}(z_1,\ldots,z_k) \\
    & (b_1,\ldots,b_k) = LSTM_{decoder}(e_1,\ldots,e_k)
    \end{aligned}
\end{equation}
where $W_{in}$ is the weight matrix with the size of ($d_{t}$ + $2d_{p}$ + $d_{o}$) $\times$ $d_{hidden}$, 
$(e_1,\ldots,e_k)$ and $(b_1,\ldots,b_k)$ are hidden states of the encoder and decoder, respectively.

\paragraph{Generating Results with Pointer Networks}

Given the encoder and decoder hidden states $(e_1,\ldots,e_k)$ and $(b_1,\ldots,b_k)$, 
we calculate the attention vector as follows:
\begin{equation}
    u_j = v^T tanh(W_1e_j + W_2b_j)
\end{equation}

\noindent
where $v$, $W_1$, and $W_2$ are learnable parameters of the model and j $\in$ (1,\ldots,k). The value of $u_j$ is used as attention weight to the $j_{th}$ input:
\begin{equation}
    p(op_j|op_1,\ldots,op_k) = softmax(u_j)
\end{equation}

\noindent
where softmax normalizes the vector $u = [u_1,\ldots,u_k]$ to be an output distribution over the inputs. 
At last, the output is a list of operation paths ranked by their distribution weights. 

It should be noted that our {\em pointer network} is a variant of the original model:  we only point to a single path from the inputs rather than generating a sequence of outputs.

\paragraph{Parameter training}
We use cross-entropy loss \cite{rubinstein1999the} to train the parameters in our model:
\begin{equation}
\label{eqa:loss}
    Loss = \sum_{y_i \in Y} - \sum_{op \in y_i} [Y_{op} \cdot log(P_{op}) + (1-Y_{op}) \cdot log(1-P_{op})]
\end{equation}
$Y_{op}$=\{1, 0\} indicates whether the operation path is an oracle path, $P_{op}$ is the outputted weight of the operation path.
We use the Adam approach \cite{kingma2014adam} in the learning process to reduce the $Loss$.

\subsection{\bf Buggy position prediction}
Once the \toolname model is trained, we can use it to predict the operation path (i.e, buggy token value + bug position + change operator) for an unseen buggy method.  We first extract AST paths towards all leaf nodes within the buggy method and assign alternatively each of the three change operators, which generates all operation paths for this method.
We then embed all these operation paths and send them into the trained model. \toolname will then return the prediction result (\ie a list where the operation paths are ranked by their output weights).

%% file: 5.setup.tex
\section{Study Design}
\label{sec:exp}

\subsection{Research Questions}
\label{sec:rqs}
\begin{itemize}[leftmargin=*]
	\item  {\bf [RQ-1] }{\em  Is \toolname effective for identifying buggy code elements in real-world programs?} We attempt the fix localization on thousands of bugs to assess the performance of \toolname. Thus, we focus on the predicted results about the buggy token without considering the associated change operator.
	\item   {\bf [RQ-2] }{\em  Does \toolname accurately predict the change operator that must be applied to fix a bug?} Given that predicting the operator alone is irrelevant, we investigate the prediction performance of \toolname for the pair of ingredients constituted by the buggy code element and the change operator. The prediction of these ingredients is  essential for accelerating both manual and automated program repair~\cite{liu2019tbar, liu2019avatar}.
	\item  {\bf [RQ-3] }{\em To what extent have our design choices influenced the performance of the neural network?} We assess the impact of splitting the tokens, encoding nodes of the AST paths, and using a fully connected layer in improving the predictive power of \toolname.
	
  \item {\bf [RQ-4]} {\em Can fine-grained fix localization help improve the precision and efficiency of patch generation?} In particular, we investigate a case study of program repair, how token-level fault localization impacts (1) patch overfitting in APR, as well as (2) the number of patch candidates that are tried before a plausible patch can be identified. 

\end{itemize}

\subsection{Subject Selection}
{\bf Patch datasets.} For training the proposed \toolname neural network-based model, we collect patches from the training dataset used by the CoCoNut~\cite{lutellier2020coconut} repair tool.
To evaluate the performance of \toolname, we collect bugs from ManySStuBs4J~\cite{karampatsis2020how} and Defects4J~\cite{just2014defects4j} datasets. 

We choose these datasets because (1) CoCoNut and ManySStuBs4J are large-scale patch benchmarks which are suitable for assessing the generalization ability of our approach and (2) Defects4J is the most widely used benchmark in the software testing literature. 
We deduplicate the samples and remove samples related to test code, since we focus on fix localization in source code. Overall, the final datasets include 436\,676, 26\,406, and 393 bugs from CoCoNut Dataset, ManySStuBs4J, and Defects4J, respectively.

{\bf Repair benchmarks.} To answer RQ4 (on the added-value of \toolname for APR patch generation performance), we consider bugs within four widely-used Java defect benchmarks: Defects4J (V1.4)~\cite{just2014defects4j}, Bears~\cite{madeiral2019bears}, QuixBugs~\cite{lin2017quixbugs} and Bugs.jar~\cite{saha2018bugs}.
We choose these benchmarks because they have been widely used for evaluating the state of the art APR tools whose assessment reports are available for comparison.


\subsection{Experiment Settings}  

\subsubsection{\bf Implementation} 
We implemented \toolname by integrating the Pytorch implementation of PointerNet\footnote{\url{https://github.com/shirgur/PointerNet}} with the framework provided for code2seq~\cite{alon2019code2seq}.
The \toolname model is trained on two servers with NVIDIA TITAN V, Xp, and 1080 Ti GPUs.

\subsubsection{\bf Parameter configuration}
Following the insights of previous studies~\cite{alon2019code2vec, alon2019code2seq}, we set a limit for the maximum \textit{length} of the AST path, as well as for the maximum number of operation paths that we feed to the neural network. These limits are respectively noted as $max_l$ and $max_k$ and are determined empirically as hyper-parameters of our model.

For tuning, we randomly sample 43\,000 patches among the considered CoCoNut patches as our validation dataset, while the remaining are used for training and testing. 
In the tuning process, we mainly focus on the key hyper-parameters: the vector length of encoded tokens (64, 128, 256), learning rate (0.001, 0.002, 0.005), epoch size (20, 40, 50), batch size (64, 128, 256), $max_l$ (10, 15, 20), and $max_k$ (100, 120, 150, 180, 200). Default values of other parameters are taken from the implementation of code2seq. The final values of hyper-parameters of our model are displayed in Table~\ref{tab:parameter-value}. 

\input{tables/parameter_value}

\subsubsection{\bf Evaluation setting}

To answer the first three research questions (c.f., Section~\ref{sec:rqs}) on the performance of the \toolname neural network-based model, we also perform 10-fold cross validation~\cite{wang2020automated,10_fold} on CoCoNut dataset to avoid bias and ensure the generalization of the model. Each training epoch takes about 8~minutes on our computing power, summing up to a total of about 5-hour execution time for each validation fold. 


\subsubsection{\bf Metrics}
To answer research questions RQ-1 to RQ-3, we use the following widely-used metrics \cite{zou2019empirical,li2019deepfl,lou2020can}:

{\bf Recall@Top-n:} Following the existing works~\cite{li2019deepfl,zou2019empirical,zhang2017boosting} for the fault localization, we selected the Recall at Top-N as the metric. 
Specifically, given a descending-order ranked list of predicted suspicious code elements $E^{sus}=\{e^{sus}_1, e^{sus}_2,...,e^{sus}_n\}$, if the buggy code element $e_i$ is within $E^{sus}$, it is localized. 
A smaller $n$ value indicates the more accuracy of fault localization for automated program repair. 
It is equivalent to the spectrum-based fault localization methods~\cite{parnin2011automated,lou2020can,li2019deepfl} that locate the bug position at the line level with a ranked list of suspicious statements.
In this work, we select $n$ as 1, 3, 5, 10, and 20. A higher {\bf Recall@Top-n} value indicates the better precision of fault localization. 
Note that in our evaluation datasets, a bug may possess more than one buggy code element (\eg a patch changes two or more code tokens). 
Following the validation procedure of spectrum-based fault localization~\cite{parnin2011automated,lou2020can}, we consider that a bug is localized at top-$n$ when one of its buggy code elements is ranked among the top-$n$ predictions of \toolname.

{\bf Mean First Rank (MFR):} It computes the mean rank value for the first localized buggy code element in the ranked list of predicted suspicious code elements.
For example, given a bug, the list predicted by \toolname contains $n$ suspicious code elements, and the buggy code element is ranked at the $k$ ($k \in [1, n]$) position in the list.
A lower {\bf MFR} value indicates the better precision of fault localization. 

The last research question RQ-4 is assessed with the following two metrics \cite{liu2020efficiency,xiong2017precise}:

{\bf Correctness Ratio (CR):} In the literature, to assess the bug fixing performance of APR tools, researchers mainly focus on the number of bugs that can be fixed by APR tools with generated plausible patches (\ie the patches can make the patched program pass all test suites) \cite{le2012genprog,kim2013automatic}.
Due to the overfitting challenge \cite{qi2015analysis,wang2020automated,tan2016anti}, the capability of generating correct patches (\ie the patches can really fix the buggy program but not just make the patched program pass all test suites) \cite{liu2020efficiency,xiong2017precise,wen2018context} is proposed to evaluate the bug fixing performance of APR tools in the community. 
Suppose that, an APR tool can generate plausible patches for $x$ bugs, and the generated patches of $y$ bugs are correct, then $CR = y/x * 100\%$.
To assess the correctness of APR-generated patches, we adopt the open patch correctness assessment rules provided by Liu \etal \cite{liu2020efficiency}, where 15 rules are explicitly defined to illustrate how to identify an APR-generated patch as correct when comparing it with the ground-truth develop-written patch provided in benchmark datasets. 
Such rules are publicly available at \url{https://github.com/TruX-DTF/APR-Efficiency} and are widely used in the patch correctness validation of APR works \cite{wang2020automated,tian2020evaluating,ye2019automated,lutellier2020coconut,jiang2021cure}.

{\bf Number of Patch Candidates (NPC):} Repair efficiency is the other metric that is used to assess the bug fixing performance of APR tools.
The time cost of generating patches has been proposed in the literature \cite{le2012systematic,ghanbari2019practical} to assess the repair efficiency.
However, time cost could be biased by differences among buggy programs, experimental setup and platforms~\cite{liu2020efficiency}. 
In this study, we leverage the NPC score, the number of patch candidates generated by an APR tool when the first plausible patch is produced~\cite{liu2020efficiency}, to assess the repair efficiency of APR tools. 

%% file: tables/parameter_value.tex
\begin{table}[!ht]
  \centering
  \caption{Hyper-parameter values inferred for \toolname. }
    {
    \begin{tabular}{cccccc}
    \toprule

{\em vector length of token} & {\em learning rate}  & {\em epoch size}  & {\em batch size} &  {\textbf{$max_l$}}  & {\textbf{$max_k$}} \\
    \hline
    128 & 0.001  & 40 & 256   & 15    & 120 \\
    \bottomrule
    \end{tabular}%
    }
  \label{tab:parameter-value}%
\end{table}%

%% file: 6.evaluation.tex
\section{Study Results}
\label{sec:eval}
We now report on the experimental results and conclude on the research questions. 

\subsection{Fix Localization Performance}


To the best of our knowledge, \toolname is the first approach that targets the fix localization of buggy code elements (cf. Section~\ref{sec:declaration}) that must be changed. 
Nevertheless, we propose to build two baseline localizers for comparison with \toolname.

{\bf Baseline\#1 -- statistics based}:  
in the study on patch granularity by Liu~\etal~\cite{liu2018closer}, a reported finding suggests that some specific code elements are more prone to be buggy than others. We leverage their data to build a simple predictor as a baseline: ranking the elements based on their probabilities of being buggy. When two or more tokens present the same probability, their rankings are further determined by their appearance orders in the token sequence of the buggy method.

{\bf Baseline\#2 -- machine learning-based}: we also propose to train a Random Forest model \cite{random_forest} for predicting buggy tokens. For a specific token, we consider four kinds of features: $\langle${\em token rank}, {\em statement type}, {\em token length}, and {\em number of sub-tokens}$\rangle$, where {\em token rank} denotes the ranking of this token in the token sequence of the buggy method, {\em statement type} denotes the type of the statement to which this token belongs (\eg {\mycode ReturnStatement} or {\mycode IfStatement}), {\em token length} denotes its number of characters.
The {\em number of sub-tokens} of each code token is the sum of sub-tokens that each code token has, after each code token is split into sub-tokens based on the camel case and underscore naming conventions.

Our experiments consider two scenarios where \toolname is provided with the full buggy method or the specific line: both inputs can indeed be provided by current spectrum-based fault localization methods. Performance results are provided in Tables~\ref{tab:rq1-line} and~\ref{tab:rq1-method} in terms of the percentages of bug cases for which the tools managed to place the buggy code element among the Top-$k$ of its ranked suspicion list. The {\em Mean First Rank (MFR)} computes the mean rank value for the correct buggy code element: the closer to 1, the better. 


\input{tables/RQ1-line}
\input{tables/RQ1-method}

The performance of Baseline\#1 is consistent with the observations in the study of Liu~\etal~\cite{liu2018closer}.
Baseline\#2 also appears to provide a bit better performance than Baseline\#1.
Overall, \toolname outperforms these baselines on all datasets and with respect to all metrics.
When the input for fix localization is a buggy method, \toolname can precisely rank the buggy code elements at the top-1 position for 46.9\% of bugs in the {\em CoCoNut} dataset, 30.7\% of bugs in the {\em ManySStuBs4J} dataset, and 34.9\% of bugs in the {\em Defects4J} dataset. In contrast, both baselines perform poorly: Baseline\#1 can localize at Top-1 only 3.7\%, 0.7\%, and 1.3\% of bugs, respectively in CoCoNut, ManySStuBs4J, and Defects4J, while Baseline\#2 localizes respectively 14.0\%, 8.1\%, 3.0\% of bugs in these datasets. Finally, we note that \toolname can achieve a recall of around 90\% in localizing the right buggy code element within its Top 20 suggestions for each bug.

\notez{{\bf [RQ-1]} \hspace{2mm}\begin{rotate}{90}\ding{242}\end{rotate} \ding{172} \toolname is largely more effective than both a statistical baseline approach and a machine learning-based baseline approach in performing fine-grained localization of buggy code elements (i.e., fix localization). }

\begin{figure*}[!ht]
    \centering
    \begin{subfigure}{.495\linewidth}
        \includegraphics[width=0.98\linewidth]{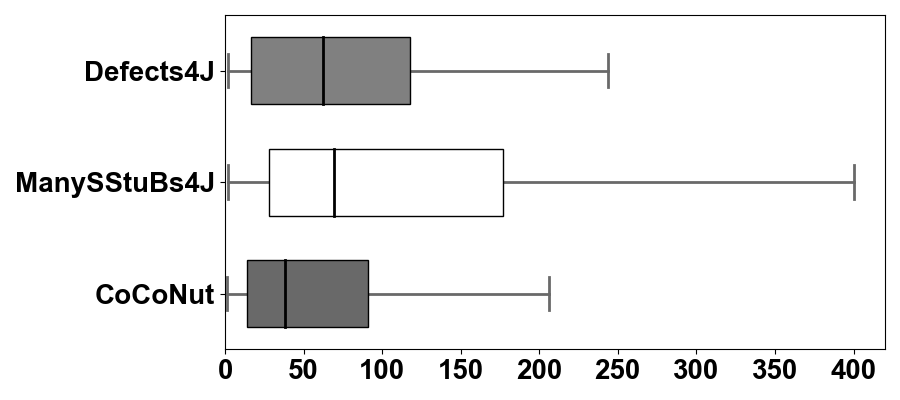}
        \caption{Buggy method}
    \end{subfigure}
    \begin{subfigure}{.495\linewidth}
        \includegraphics[width=0.98\linewidth]{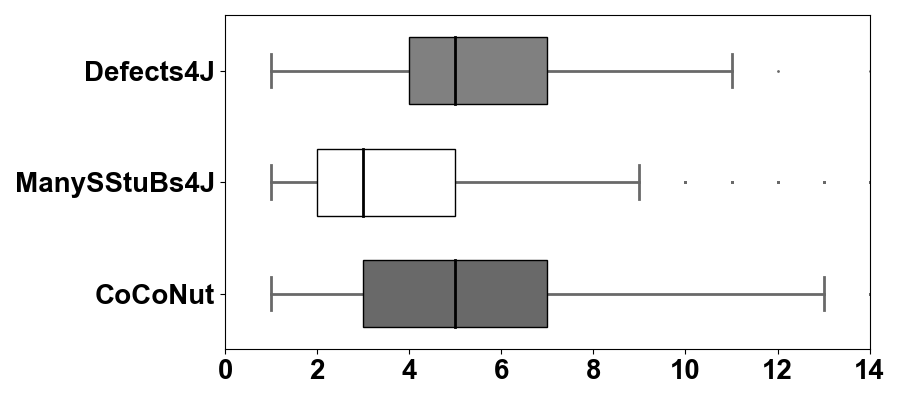}
        \caption{Buggy line}
    \end{subfigure}
    \caption{Number of tokens in each buggy method and buggy line.}
    \label{fig:number_of_tokens}
\end{figure*}

We further investigate how much manual debugging effort can  be saved when using \toolname to localize the tokens that must be changed in a method.
To that end, we count the number of code tokens in each buggy method and buggy line identified from the three defects benchmarks used in this study. 
We exclude the Java keywords (\eg {\em if}, {\em int}, etc.) when calculating the results since it is unlikely that these tokens are buggy. This configuration, however, may under-estimate the effectiveness of \toolname since there indeed exists bugs whose fixes require the changes of such keywords (\eg Math-57 from Defects4J benchmark is patched by changing {\mycode int} into {\mycode float}).
Results are shown in Fig.~\ref{fig:number_of_tokens}.

We note that the medium values of the number of tokens in each buggy method are 38, 72, and 67 for the CoCoNut, ManySStuBs4J, and Defects4J benchmarks, respectively. 
Recall that the Mean First Rank (MFR) values for \toolname on these three benchmarks are 3.9, 6.7, and 6.5 respectively (cf. Table~\ref{tab:rq1-method}).
Such results indicate that \toolname can rank the buggy elements at a rather high position (\ie usually in the top 10\% among all the method code elements). When it comes to the number of tokens in each buggy line, the medium values become 5, 3, and 5 for the three benchmarks.
From Table~\ref{tab:rq1-line}, the MFR values for \toolname are 2.3, 1.3, and 1.4 respectively, which means \toolname can generally rank the buggy elements within the first half.

\notez{{\bf [RQ-1]} \hspace{2mm}\begin{rotate}{90}\ding{242}\end{rotate} \ding{173}
When provided with a buggy method, \toolname can drastically reduce debugging effort needed by developers to pinpoint the buggy tokens: on average, \toolname filters out about 90\% of non-buggy code tokens. Even in such a case where the buggy line is known (e.g., with spectrum-based fault localization), \toolname can still halve the number of tokens to manually check.
}

\noindent
\subsection{Joint Prediction of Bug and Change Operator}
Aside from the question of accurate fine-grained bug localization, manual repair as well as APR are challenged in the selection of adequate repair operators. Classically, several approaches in the APR literature rely on heuristics to try-and-err with different operators~\cite{jiang2018shaping,liu2019tbar,wen2018context}.

\toolname performs the selection of the change operators for the identified buggy code elements by learning a model from the operation paths (i.e., AST path + operator) of existing patches.
The prediction performance of \toolname (given a buggy method as input) is detailed in Table~\ref{table:RQ2-result}. 

\input{tables/RQ2-result}

We note that the performance of joint prediction is decreased only slightly when compared against the prediction of buggy code elements alone (cf. Table~\ref{tab:rq1-method}). This suggests that the \toolname model is able to predict the change operator when it precisely locates the buggy element.


\notez{{\bf [RQ-2]} \hspace{2mm}\begin{rotate}{90}\ding{242}\end{rotate} We confirm that \toolname can not only localize the buggy code element (RQ1) but also accurately predict the code change operator associated to it.}

\subsection{Ablation Study}
We assess the contributions of some components and design choices in the performance of our core prediction model, \toolname. Our analysis focuses on three key aspects: (1) the impact of our sub-token splitting; (2) the impact of our node-based AST path encoder; and (3) the impact of fully-connected layer for feature extraction.

In the first experiment assessing the impact of sub-token splitting, we directly feed each token in our dataset to $E_t(*)$. 
In the second experiment assessing the design choice of considering each node, we do not represent each node in the AST path $p$. Instead, we learn an embedding matrix $E_{paths}$ for representing the whole AST path.
In the third experiment, we drop the fully-connected layer: we directly use the concatenated operation path representation as the input of the LSTM encoder layer. 

The results are summarized in Table~\ref{tab:ablation_onlyMFR} w.r.t. the performance of the bug prediction model. Due to space constraints, we only provide MFR metric values which are indicative of the overall performance.
We note that under each experimental setting, the MFR value when {\em no token splitting} is applied is always the highest.
On CoCoNut dataset, while the performance of model variants can be on par with that of the full \toolname, the MFR value is always lowest with \toolname, indicating that our design choices globally lead to the best predictive model.
On ManySStuBs4J and Defects4J datasets, the results of \toolname are systematically better than those yielded by its variants. Note that the different models are trained on the CoCoNut dataset, which, for applied on ManySStuBs4J and Defects4J bugs, leave room to improve the neural network design choices for \toolname.

\input{tables/ablation_onlyMFR}

\notez{{\bf [RQ-3]} \hspace{2mm}\begin{rotate}{90}\ding{242}\end{rotate} Various design choices have together contributed to the performance of \toolname. Splitting the tokens before embedding appear to be the most rewarding design choice.}

\subsection{Repair Performance on Real Bugs}

In this RQ, we seek to investigate if narrowing the mutation space of APR tools by leading them to change the predicted buggy tokens can enhance the precision (\ie avoiding generating overfitting patches) and the efficiency (\ie reducing the number of patch candidates that are tried before the first plausible patch). 
After a review of literature artifacts, we found that state-of-the-art APR tools are implemented such that the patch generation process is entangled with current fault localization settings.
We thus propose to experiment with implementations of two straightforward repair pipelines based on code completion and on heuristics. The basic workflow for these two pipelines is as follows:

\noindent
{\bf Input:} We assume that the faulty method is known. This is a reasonable assumption since, based on data provided in recent studies \cite{lou2020can,benton2020effectiveness}, existing techniques are effective in localizing bugs at the method level granularity. Some APR tools~\cite{liu2019you,le2016history} in the literature are even assessed based on the assumption.

\noindent
{\bf Bug prediction:} We use \toolname to predict the buggy tokens. After this step, we obtain a ranked list of operation paths.

\noindent
{\bf Operation path selection:} We consider the first 20 operation paths since our evaluation results (cf. Table~\ref{table:RQ2-result}) have shown that \toolname has a very high recall within its top-20 results (i.e., the searched operation path is in among those).
We then use a simple try-and-err heuristic: we iterate over each path following the ranking order and input it to the patch generation process. If all paths have been tried but no test-adequate patch is generated, we re-iterate over pairwise combinations of operation paths, and so on.
This process comes to an end when a patch that passes all the tests is generated.

\noindent
{\bf Patch generation:} In this step, patches are generated by using either a code completion technique or straightforward heuristics. We will introduce the details of these two methods later in this section (after the general workflow).

\noindent
{\bf Patch validation:} Each generated patch is then applied to the program, on which the test suite is executed. A valid patch must make the program pass all the test cases. When such a patch is found, the search stops. We then manually (but systematically) check whether the generated plausible patch is indeed correct. Correctness is assessed in comparison with the ground truth patch. We adopt the rules that were presented by Liu \etal \cite{liu2020efficiency} to ensure that our assessment is replicable.

\subsubsection{Code completion for patch generation}
Given the naturalness of software~\cite{hindle2012naturalness}, we postulate that it should be possible to statically predict missing elements just like one predicts missing words in natural language text. Fortunately, such an idea has been very recently explored by Alon \etal with their approach for code generation with structural language modeling (SLM)~\cite{alon2020structural}. 
While other state of the art such as seq2seq~\cite{sutskever2014sequence}, code2seq~\cite{alon2019code2seq} and structured prediction~\cite{brockschmidt2018generative} are valid alternatives for our code completion task, we build in this work on SLM: its {\em AnyCodeGen} trained model was experimentally shown to significantly outperform a variety of strong baselines on Java and C\# languages for the task of code completion~\cite{alon2020structural}.

Once \toolname yields the predicted operation path, we parse the change operator and the token (buggy element). If the operator is {\tt DELETE}, then we simply generate a patch that removes this token from the method at the position identified by the predicted AST path. If the operator is {\tt UPDATE}, however, we create a hole (represented by the ``??'' notation\footnote{This notation is recognized by \ACG as a program hole to be completed.}) within the buggy method at the position of the identified token for the predicted AST path. The holed code thus constitutes an input for querying the \ACG structural language model-based engine. Finally, if 
the change operator is {\tt INSERT}, we simply add a placeholder (still represented by the ``??'' notation) to the method, again forming a query to the code completion engine.

\subsubsection{Heuristics-based repair pipeline}
In this implementation, we generate patches via using the following heuristics:
\begin{itemize}[leftmargin=*]
\item If the predicted operation is {\tt DELETE}, we directly remove this token. 
\item If the predicted operation is {\tt INSERT} or {\tt UPDATE}, we transform the program using the following rules:
\begin{itemize}[leftmargin=*]
\item if the token is an operator, we change it to another one in the operator set (\eg == $\rightarrow$ != and $>=$ $\rightarrow$ $>$);
\item if the token is a boolean value, we change it to its reverse (\eg false $\rightarrow$ true);
\item if the token is a data type, we change it to another type (\eg int $\rightarrow$ float);
\item if the token is an identifier, we replace it or pad the location with its 5-nearest\footnote{Distance is computed within the token sequence of the buggy method.} identifiers with the same type. We consider generating only 5 candidates to remain comparable to  \ACG code completion engine which only returns 5 code completion results.
\end{itemize}
\end{itemize}

\noindent
{\bf Analysis of repairability.}
We provide in Table~\ref{tab:repair-result-simple} the performance of our repair pipelines on the four defect benchmarks. We also compare their performance against the results of 26 APR techniques reported in recent literature~\cite{jiang2021cure}. Due to space limitation, we only include in this table the most representative tools w.r.t. yielding highest precision (i.e., high correctness ratio - CR - above part of the table) and w.r.t. repairability (i.e., the highest number of bugs fixed - below part of the table).

Overall, we find that our pipelines can both achieve 100\% precision: all generated patches that passed the test suite were found to be correct. Actually, during the manual check, we found that they were all identical to the ground-truth patches. In comparison, the APR tool with the highest precision in the literature, i.e., CapGen, has a correctness ratio of 84\%. We also note that our pipeline provides comparable performance with respect to the number of correctly fixed bugs when compared with the first four tools. For instance, ACS can only correctly fix 18 bugs while \toolname (with heuristics-based patch generation) can correctly fix 21 bugs.
On the other hand, although tools with high repairability metrics outperform our pipelines in terms of the number of correctly fixed bugs, their overall precision is low. For example, DLFix plausible patches even include more incorrect ones than correct ones. Since in practice developers should check that the correctness of generated patches before integrating them, it may be detrimental to use APR if the correctness ratio is low.

We also note that when using simple heuristics for patch generation, we can fix more bugs (32 vs. 27) than when relying on a carefully-designed and fully-trained sophisticated Deep Learning model for code completion (\ie \ACG). Our results suggest that, for program repair, finding donor code within the same method may be more effective than predicting missing tokens using big data.
Moreover, we found that our pipelines are able to fix bugs which were never yet reported to be fixed by the considered 26 APR tools. 2 of such bugs can be found in the Defects4J dataset. \toolname + code completion also led to fixing 2 bugs from Bears as well as  4 bugs from Bugs.jar which were not fixed by the state-of-the-art tools (such as ARJA~\cite{yuan2018arja}) which were applied on these benchmark. These results further demonstrate token-level-based program repair is a sweet spot for developing further directions in generate-and-validate APR.

\input{tables/repair_result_simple}

\notez{{\bf [RQ-4]} \hspace{2mm}\begin{rotate}{90}\ding{242}\end{rotate} \ding{172} Automated program repair built on the top of \toolname localization results can produce patches with high correctness ratio, hence providing a novel perspective to address patch overfitting in program repair.
}

\begin{figure}[!ht]
	\centering
	\includegraphics[width=0.7\linewidth]{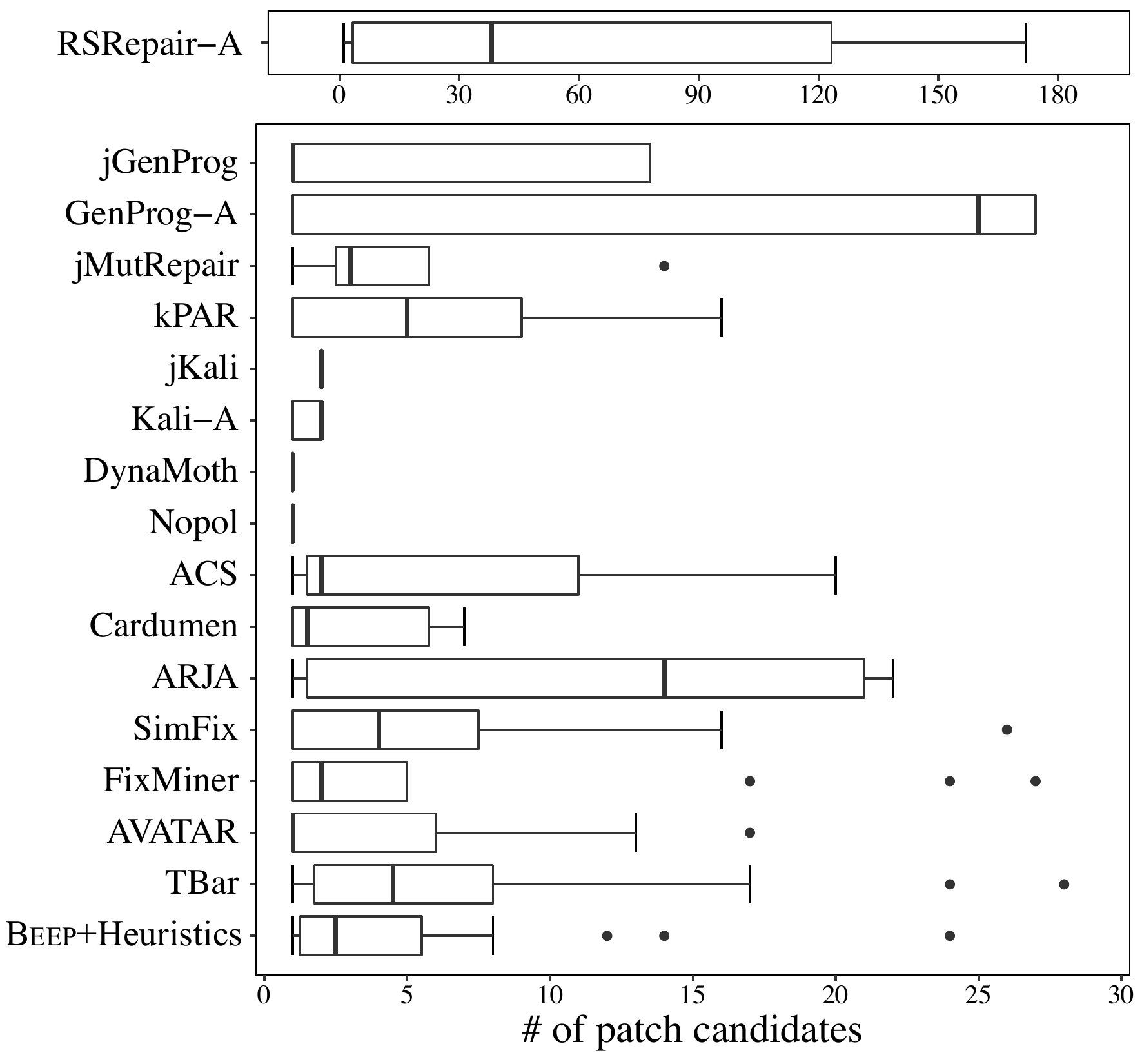}
	\caption{Comparison of efficiency.}
	\label{fig:efficiency}
\end{figure}

\noindent
{\bf Analysis of patch generation efficiency.}
In this section, we consider only the \toolname plus heuristics repair pipeline since it is more effective.
Repair efficiency assessment considers the number of patch candidates (NPC) that are generated and tested before a plausible patch is hit. NPC score was recently proposed by Liu~\etal~\cite{liu2020efficiency} to enable fair comparison across computing environments. They provide the NPC score of several state-of-the-art tools for the Defects4J bugs with the assumption that each APR is provided with the buggy line. We also place the proposed pipeline under this setting. Fig.~\ref{fig:efficiency} compares the distribution of NPC of the pipeline vs. 16 APR tools considered by Liu~\etal~\cite{liu2020efficiency} on Defects4J bugs.

With a median NPC score of 2, our pipeline is more efficient than pattern-based APR techniques (e.g., TBar and kPAR) and even more recent state of the art heuristic-based APR techniques (e.g., SimFix). Nevertheless, six tools significantly outperform this pipeline in terms of efficiency: three (i.e., DynaMoth, Nopol, ACS) actually use constraint-solving and synthesis strategies, which indeed make them efficient (see ~\cite{liu2020efficiency});  the other three (i.e., jKali, Kali-A, and jMutRepair) apply naive repair strategies that limit their search spaces -- jKali and Kali-A change conditional statements into {\em true} or {\em false} and jMutRepair only mutate operators.

\notez{{\bf [RQ-4]} \hspace{2mm}\begin{rotate}{90}\ding{242}\end{rotate} \ding{173} Successful prediction of the buggy code element and adequate change operator by \toolname limits the number of patch candidates, hence improving the efficiency of patch generation.}

%% file: tables/RQ1-line.tex
\begin{table}[!t]
  \centering
  \caption{Fix localization performance - {\em buggy line} as input.}
  {
  \begin{threeparttable}
    \begin{tabular}{l|c|C{15mm}C{15mm}C{15mm}|C{15mm}}
    	\toprule
    	Dataset & Tool & Top-1 & Top-3 & Top-5 & MFR \\
    	\hline
    	\multirow{2}[4]{*}{CoCoNut} 
    	  	& Baseline\#1 & 26.1\% & 65.9\% & 86.7\% & 3.2 \\\cline{2-6}
    	  	& Baseline\#2 & 61.7\% & 76.5\% & 89.9\% & 2.7 \\\cline{2-6}
    		&\cellcolor{black!25}\toolname & \cellcolor{black!25}63.8\% & \cellcolor{black!25}83.9\% & \cellcolor{black!25}91.2\% & \cellcolor{black!25}2.3 \\
    	\hline
    	\multirow{2}[4]{*}{ManySStuBs4J} 
    		& Baseline\#1 & 36.7\% & 84.1\% & 94.0\% & 2.5 \\\cline{2-6}
    		& Baseline\#2 & 36.9\% & 86.3\% & 97.2\% & 1.7  \\\cline{2-6}
    		&\cellcolor{black!25}\toolname & \cellcolor{black!25} 85.2\% & \cellcolor{black!25}96.3\% & \cellcolor{black!25}100\%  & \cellcolor{black!25}1.3 \\
    	\hline
    	\multirow{2}[4]{*}{Defects4J}
    	    & Baseline\#1 & 18.2\% & 49.4\% & 83.1\% & 2.8 \\\cline{2-6}
    	    & Baseline\#2 & 23.9\% & 61.8\% & 86.7\% & 1.9 \\\cline{2-6}
    	    &\cellcolor{black!25}\toolname & \cellcolor{black!25} 86.7\% & \cellcolor{black!25} 91.4\%  & \cellcolor{black!25} 96.8\%  & \cellcolor{black!25} 1.4 \\
    	\bottomrule
    \end{tabular}%
    {\footnotesize CoCoNut Dataset: 436\,676 bugs, ManySStuBs4J: 26\,406 bugs, and Defects4J: 393 bugs, the same as Table~\ref{tab:rq1-method}.}
  \end{threeparttable}
   }
  \label{tab:rq1-line}%
\end{table}%

%% file: tables/RQ1-method.tex
\begin{table}[!t]
  \centering
  \caption{Fix localization performance - {\em buggy method} as input.}
  {
    \begin{tabular}{l|c|C{12mm}C{12mm}C{12mm}C{11mm}|C{10mm}}
    	\toprule
    	Dataset & Tool & Top-1 & Top-5 & Top-10 & Top-20 & MFR \\
    	\hline
    	\multirow{2}[4]{*}{CoCoNut} 
    		& Baseline\#1 & 3.7\%  & 25.3\% & 40.9\% & 57.9\% & 38.7 \\\cline{2-7}
    		& Baseline\#2 & 14.0\% & 31.9\% & 41.9\% & 48.3\% & 32.1 \\ \cline{2-7}
    		&\cellcolor{black!25}\toolname & \cellcolor{black!25}46.9\% & \cellcolor{black!25}74.2\% & \cellcolor{black!25}85.5\% & \cellcolor{black!25}95.2\% & \cellcolor{black!25}3.9 \\
    	\hline
    	\multirow{2}[4]{*}{ManySStuBs4J} 
    		& Baseline\#1 & 0.7\%  & 10.4\% & 22.3\% & 40.5\% & 42.0 \\\cline{2-7}
    		& Baseline\#2 & 8.1\% & 20.9\% & 33.6\% & 46.2\% & 28.9 \\\cline{2-7}
   			&\cellcolor{black!25}\toolname & \cellcolor{black!25}30.7\% & \cellcolor{black!25}56.4\% & \cellcolor{black!25}72.6\% & \cellcolor{black!25}90.1\% & \cellcolor{black!25}6.7 \\
   		\hline
   		\multirow{2}[4]{*}{Defects4J} 
    		& Baseline\#1 & 1.3\%  & 13.3\% & 25.3\% & 45.3\%  & 65.1 \\\cline{2-7}
    		& Baseline\#2 & 3.0\% & 23.9\% & 29.8\% & 31.3\% & 32.1 \\\cline{2-7}
   			&\cellcolor{black!25}\toolname & \cellcolor{black!25} 34.9\% & \cellcolor{black!25} 57.1\% & \cellcolor{black!25} 68.2\%& \cellcolor{black!25} 87.3\%& \cellcolor{black!25} 6.5\\
    	\bottomrule
    \end{tabular}%
  }
  \label{tab:rq1-method}%
\end{table}%

%% file: tables/RQ2-result.tex
\begin{table}[!ht]
  \centering
  \caption{Joint prediction performance: buggy code element {+ \bf code change operator} with the buggy method as input.}
  {
    \begin{tabular}{l|cccc|r}
    \toprule
    Dataset & Top-1 & Top-5 & Top-10 & Top-20 & MFR \\
    \midrule
    CoCoNut  & 44.6\% & 67.8\% & 77.9\% & 87.5\% & 7.5 \\
    ManySStuBs4J   & 29.7\% & 50.8\% & 63.7\% & 79.3\% & 11.5 \\
    Defects4J & 29.2\% & 51.8\% & 57.7\% & 65.5\% & 12.3 \\
    \bottomrule
    \end{tabular}%
    }
    \label{table:RQ2-result}
\end{table}%

%% file: tables/ablation_onlyMFR.tex
\begin{table}[!t]
  \centering
  \caption{Comparison of Model Designs.}
  {
    \begin{tabular}{l|r|c|c|c}
    \toprule
    \multirow{2}{*}{\makecell[l]{{\bf Prediction}\\{\bf objective}}} & \multirow{2}{*}{{\bf Model Design}} & \multicolumn{3}{c}{\bf MFR} \\\cline{3-5}
    & & {\bf CoCoNut} & {\bf ManySStuBs4J} & {\bf Defects4J} \\
    \hline
    \multirow{4}[1]{*}{\makecell[l]{Buggy code\\ element}} & no token splitting &  4.7  &  7.2 &  7.3\\
          & no AST nodes &  {\bf 3.9}   & 7.2 & 7.2 \\
          & no fully-connected layer &  4.0  &  7.0 & 6.7 \\
          & \toolname   &  {\bf 3.9}  & {\bf6.7} & \bf{6.5} \\\hline
    \multirow{4}[2]{*}{\makecell[l]{Element +\\ Operator}} & no token splitting &  13.5 & 16.4 & 16.4\\
          & no AST nodes  & 7.6  &  12.3 & 14.6\\
          & no fully-connected layer &  8.3  & 12.6 & 13.7 \\
          & \toolname   &  {\bf7.5}  &  {\bf11.5} & \bf{12.3} \\
    \bottomrule
    \end{tabular}%
    }
  \label{tab:ablation_onlyMFR}%
\end{table}%

%% file: tables/repair_result_simple.tex

\begin{table}[!t]
  \centering
  \caption{Comparison with state-of-the-art APR tools on fixing real-world bugs.}
  {
  \begin{threeparttable}
    \begin{tabular}{lC{18mm}C{18mm}C{18mm}C{16mm}R{12mm}}
    \toprule
    {\bf APR Tool} & {\bf Defects4J} & {\bf Bears} & {\bf QuixBugs} & {\bf Bugs.jar} & {\bf CR(\%)} \\
    \hline
    JAID~\cite{chen2017contract}  & 25/31   & -     & - &-    & 80.6\% \\
    CapGen~\cite{wen2018context} & 21/25 & -     & - &-     & 84,0\% \\
    ACS~\cite{xiong2017precise}   & 18/23   & -     & -  & -   & 78.3\% \\
    FixMiner~\cite{koyuncu2020fixminer} & 25/31 & - & - & - & 80.6\% \\
    \hline
    SimFix~\cite{jiang2018shaping} & 34/56 & - & - & - & 60.7\% \\
    DLFix~\cite{li2020dlfix} & 30/65 & -  & - & - & 46.2\% \\ 
    TBar~\cite{liu2019tbar}  & 43/81 & -     & -  & -   & 53.1\% \\
    CURE~\cite{jiang2021cure} & 57/104 & - & 26/35 & - & 59.7\% \\
    \hline
    \rowcolor{grey}\toolname + Code completion & 16/16 (2) & 2/2 (2) & 4/4 (0) & 5/5 (4) & \textbf{100\%} \\
    \hline
    \rowcolor{grey}\toolname + Heuristics & 21/21 (2) & 0/0 (0) & 5/5 (0) & 6/6 (4) & \textbf{100\%} \\
    \bottomrule
    \end{tabular}%
    {\footnotesize Data of other tools are extracted from the original papers. `-' denotes no relevant data. x/y represents the tool generates y plausible patches among which x are correct on this benchmark.
    Numbers in the parentheses denote the numbers of bugs which are not fixed by the state-of-the-art before.
    }
    \end{threeparttable}
    }
  \label{tab:repair-result-simple}%
\end{table}%

%% file: 7.discussion.tex
\section{Discussion}
\label{sec:dis}

\noindent{\bf Influence of bug occurrence:}
\toolname, as similar data-driven approaches~\cite{li2020dlfix}, performs poorly on infrequent bugs that are not common in programs or their similar bugs are very rarely occurred. Indeed, features of such bugs are generally not well captured in model training.

\noindent{\bf Declaration vs body of methods:} Buggy method declarations are challenging for \toolname. For example, in Defects4J bug {\em Lang-29}, the return type of the method should be {\em float} instead of {\em int}. Given that such bugs have short AST paths (2 nodes), the model fails to learn any relevant features for predicting the buggy elements.

\noindent{\bf The ability to fix multi-location bugs:}
Despite that, the majority of bugs which can be fixed by \toolname with the heuristics are about the modification of a single code token, we did note that this pipeline can fix multi-location bugs. Totally, our pipeline fix two multi-location bugs from Defects4J dataset.
A case is illustrated in Figure~\ref{fig:math-79}. 
Fixing this bug requires to change the data type {\tt int} into {\tt double} in two locations. After \toolname accurately predicts the two buggy elements (and the operators) within its Top 2, our heuristics successfully generate the correct patch since it allows to change the data type.   
Another instance is Time-4 where our pipeline precisely predicts the deletion of a token with an insertion in another location. 

\begin{figure}[!ht]
	\centering
 	\vspace{3mm}
\begin{lstlisting}[language=diff, firstnumber=1, numbers = left,xleftmargin=2em, frame=lines,]
// Ground-truth patch for Math-79

 public static double distance(int[] p1, int[] p2) {
-       int sum = 0;
+       double sum = 0;
        for (int i = 0; i < p1.length; i++) {
-           final int dp = p1[i] - p2[i];
+           final double dp = p1[i] - p2[i];
            sum += dp * dp;
        }
        return Math.sqrt(sum);
    }
\end{lstlisting}
	\caption{The ground-truth patch for the bug Math-79.}
  \label{fig:math-79}
\end{figure}

\noindent{\bf Threats to Validity:}
Our experimental results carry some threats to validity, notably \ding{182} its generalization beyond the Java programming language. Although the theoretical design of \toolname is valid for any language whose AST can be readily generated (e.g., C/C++), we selected Java due to the availability of large-scale off-the-shelf datasets and benchmarks, and the possibility to compare against some open released state of the art tools.
As a second threat, we note that \ding{183} \toolname was trained and evaluated with two recent large-scale datasets with a significant proportion of single-statement patches. Such a dataset may be biased or non-representative. We plan to extend the evaluations to more kinds of bugs by collecting more diverse datasets.
As a third threat, \ding{184} Defects4J contains bug cases where the fix consists in adding or deleting whole statements. The localization of such bugs is a good target for spectrum-based fault localization, but not of \toolname because we did not consider such cases in our training set (borrowed from CoCoNut).
As a forth threat, \ding{185} \toolname is built on an assumption of the available information of fault methods/statements, that is not practical. \toolname is a token-level fault localization built on top of the statement-level/method-level fault localization tools. The assumption is mainly used to evaluate the possibility of predicting the exact buggy code elements for program repair within a limited search space of suspicious bug locations. 
As a fifth threat, \ding{186} the AST/Operation paths created by \toolname do not have long length, which arises a threat for \toolname that the impact from the long path on the learning model cannot be assessed. We list it as a part of future work for improving \toolname.
Finally, \ding{187} different embeddings of the sub-tokens, AST path, and operations can  impact the performance of the deep learning models, which could further influence the performance of \toolname, nonetheless, this work mainly focuses on exploring the possibility of fine-grained fix localization for program repair. The influence of the different embeddings will be investigated in future work.

%% file: 8.relatedwork.tex
\section{Related Work}
\label{sec:relatedWork}



\noindent{\bf Developers' Opinion on Fault Localization.}
A number of studies investigate developers' perspectives on current FL techniques. Parnin and Orso~\cite{parnin2011automated} find that several assumptions made by automated debugging techniques do not hold in practice for developers (\eg ``examining a faulty statement in isolation is enough for a developer''). Xie~\etal~\cite{xie2016revisit} find that simply providing the ranked suspicious code lines for developers actually reduces their debugging efficiency. Kochhar \etal~\cite{kochhar2016practitioners} find that the most popular FL granularity so far (\ie method level) only gains preferences from around half of the investigated practitioners. 
Our work provides developers with a finer-grained FL results by targeting the buggy code tokens as well as the required change operator. We expect \toolname research direction to contribute to reducing the burden of manual debugging in practice.

\vspace{1mm}
\noindent{\bf Learning-Based Fault Localization.}
With the power of advanced neural networks, a number of works have been proposed for performing fault localization based on machine/deep learning.
Wong and Qi \cite{wong2009bp} propose a fault localization approach based on the back-propagation (BP) neural network.
MULTRIC~\cite{xuan2014learning} is the first learning-to-rank fault localization technique that can integrate the suspiciousness values calculated with spectrum-based fault localization techniques (SBFL) to improve the accuracy of SBFL~\cite{zou2019empirical}.
Since then, program invariant~\cite{b2016learning} and source code complexity information~\cite{sohn2017fluccs} have been explored to combine the SBFL suspiciousness values for more effective learning-to-rank fault localization. 
These learning-to-rank fault localization techniques aim to rank faulty statements higher than correct ones, while other learning based fault localization techniques focus on the test coverage information~\cite{briand2007using,zhang2017deep,zheng2016fault, wong2009bp} that cannot distinguish elements accidentally executed by failed tests and the actual faulty elements~\cite{li2017transforming}.
Zhang \etal \cite{zhang2019cnn} construct a convolutional neural network customized for fault localization.
Li \etal \cite{li2019deepfl} propose to leverage the traditional MLP and RNN networks to learn the effective existing/latent features (collected from the various suspiciousness-value-based, fault-proneness-based and textual-similarity-based features from the fault localization, defect prediction and information retrieval areas) for precise fault localization.
Li \etal \cite{li2021fault} treat FL as an image pattern recognition problem and achieve so via novel code coverage representation learning (RL) and data dependencies RL for program statements.
These works all locate the buggy code at the statement and method levels which is still coarse-grained, our work explores to leverage the deep learning techniques to predict the buggy code elements in buggy statements for program repair.

\vspace{1mm}
\noindent{\bf Fault Localization for APR.}
An essential step in APR is identifying the locations of buggy code within the program to be repaired~\cite{bohme2017where}. These locations are the target for selecting code entities that must be transformed to generate patches. 
As reported in recent work~\cite{liu2019you}, the commonly-adopted FL configuration (with GZoltar and Ochiai) still provides too inaccurate bug localization information for automated program repair.
Therefore, researchers try to enhance FL for APR with predicate switching~\cite{xiong2017precise,zhang2006locating}, test case purification~\cite{jiang2018shaping}, deep learning~\cite{li2019deepfl} and other information~\cite{zhang2017boosting,wen2019historical}. 
Most APR tools leverage localization information at the granularity of code lines. To the best of our knowledge, our work is the first to target the identification of buggy code elements nested within buggy lines, thus providing a more fine-grained localization for APR. 
As reported by previous works \cite{liu2020efficiency, liu2019you}, the accuracy of FL results can cause significant differences in repair performances. In detail, accurate FL can enhance the efficiency as well as the precision of APR tools.
Our evaluation with a fine-grained token-level FL also demonstrates this point.



\vspace{1mm}
\noindent{\bf APR for Fault Localization.}
Previously, the connection between the two key points in software debugging (\ie fault localization and program repair) is that program repair techniques usually use off-the-shelf fault localization techniques to identify potential buggy locations for patching \cite{wen2018context,jiang2018shaping}. Recently, however, Lou \etal \cite{lou2020can} propose to utilize the patch execution information during program repair for providing fault localization with useful feedback. A follow-up work~\cite{benton2020effectiveness} further demonstrated that the effectiveness of this approach can be enhanced when integrating patch generation information from diverse APR systems.
Nevertheless, these works target accurate localization at the method level. They can therefore be complemented by \toolname to achieve fine-grained FL (\ie fix localization).

\vspace{1mm}
\noindent{\bf State-of-the-art APR.}
The APR community has explored various APR techniques to address the bugs in different language-written programs.
One of the commonly studied approaches is generate-and-validate program repair. 
Since GenProg~\cite{weimer2009automatically} created the milestone of program repair, which was proposed to solve C program bugs with a generic method, various APR techniques have been proposed in the literature that can be categorized into four main categories: heuristic-based, constraint-based, and template-based and learning based repair approaches~\cite{le2019automated}.
Heuristic-based repair approaches construct and iterate over a search space of syntactic program modifications, e.g., SimFix~\cite{jiang2018shaping}.
Constraint-based repair approaches generally leverage the specific technique (e.g., constraint solving) to infer the constraints of the given buggy program, which are further used to guide the generation and validation of patches, e.g., SemFix~\cite{nguyen2013semfix}.
Template-based repair approaches utilize the fix patterns identified from the real-world patches, that can provide the dedicated code change behaviors for bug fixing, e.g., TBar~\cite{liu2019tbar}.
Learning-based repair approaches rely on machine learning or deep learning techniques to learn correct fix ingredients for patch generation, e.g., Prophet~\cite{long2016automatic}.
To date, the state-of-the-art APR research work mainly focuses on the bugs in C/C++ programs~\cite{nguyen2013semfix, mechtaev2016angelix, tan2015relifix, long2015staged, long2016automatic, gao2019crash} and Java programs~\cite{kim2013automatic}.
This work is to provide the fine-grained fault localization for APR to reduce the possibility of generating plausible but incorrect patches because of the coarse fault localization.

\vspace{1mm}
\noindent{\bf Deep Learning for APR.}
A considerable number of studies have utilized deep learning techniques for program repair.\\
\noindent
\ding{228} SequenceR~\cite{chen2019sequencer}, DLFix~\cite{li2020dlfix}, CoCoNut~\cite{lutellier2020coconut}, and CURE~\cite{jiang2021cure} work at the line level, thus requiring as input the exact buggy statement, an information granularity that current fault localization inaccurately provides. In contrast, our repair pipeline only targets buggy methods, a granularity that is more accessible to off-the-shelf fault localization tools~\cite{lou2020can,benton2020effectiveness}. \\ 
\noindent
\ding{228} Although the approach of Tufano~\etal~\cite{tufano2019empirical} works with methods, they have to disregard methods with more than 50 tokens due to the performance limitation of their seq2seq translation. In contrast, we accept any-size methods.\\
\noindent
\ding{228} Allamanis~\etal~\cite{allamanis2018learning} use graph-based deep learning while Vasic~\etal~\cite{vasic2019neural} adopt multi-headed pointer networks for repairing bugs. Hellendoorn~\etal~\cite{hellendoorn2020global} further propose a hybrid model by analyzing the weakness of the previous two works. The limitation of these works is that they only focus on variable-misuse bugs. While they propose to localize variable tokens, our work targets any buggy token (and its associated change operator) within a method.\\
\noindent
\ding{228} Some other approaches~\cite{gupta2017deepfix,mesbah2019deepdelta,santos2017finding,santos2018syntax,bhatia2018neurosymbolic} focus on learning to fix defects related to the programming language syntax, some of which only apply to small-scale student programs \cite{santos2018syntax,bhatia2018neurosymbolic}. In contrast, our work targets semantic defects in large-scale real-world programs.\\
\noindent
\ding{228} Phoenix~\cite{bavishi2019phoenix} also synthesizes repairs for bugs. However, it uses static analysis as an oracle while we rely on test cases.

%% file: 9.conclusion.tex
\section{Conclusion}
\label{sec:conc}
We tackle the ambition of implementing fix localization as the fine-grained identification of buggy code elements. Our \toolname neural network-based architecture can accurately predict buggy tokens and the required change operator.
It can thus save great efforts for manual debugging.
We also assessed two intuitive repair pipelines with \toolname, where the identified buggy code elements are considered as missing and, thus, must be replaced via using either a code completion engine or a set of code search heuristics. 
We explore the bug fixing performance of the proposed pipelines on bugs across several benchmarks.
Results reveal that all patches generated by the two pipelines are found to be correct. Hence, program repair built on top of fine-grained fix localization offers a promising perspective to address the long-standing challenge of overfitting in automated program repair.

\noindent
{\bf Artefacts:} All data and source code in the study are publicly available at:
\begin{center}
{\bf \url{http://doi.org/10.5281/zenodo.4717352}}.
\end{center}

%% file: main.bbl

\begin{thebibliography}{108}


\ifx \showCODEN    \undefined \def \showCODEN     #1{\unskip}     \fi
\ifx \showDOI      \undefined \def \showDOI       #1{#1}\fi
\ifx \showISBNx    \undefined \def \showISBNx     #1{\unskip}     \fi
\ifx \showISBNxiii \undefined \def \showISBNxiii  #1{\unskip}     \fi
\ifx \showISSN     \undefined \def \showISSN      #1{\unskip}     \fi
\ifx \showLCCN     \undefined \def \showLCCN      #1{\unskip}     \fi
\ifx \shownote     \undefined \def \shownote      #1{#1}          \fi
\ifx \showarticletitle \undefined \def \showarticletitle #1{#1}   \fi
\ifx \showURL      \undefined \def \showURL       {\relax}        \fi
\providecommand\bibfield[2]{#2}
\providecommand\bibinfo[2]{#2}
\providecommand\natexlab[1]{#1}
\providecommand\showeprint[2][]{arXiv:#2}

\bibitem[\protect\citeauthoryear{Abreu, Zoeteweij, and Van~Gemund}{Abreu
  et~al\mbox{.}}{2007}]%
        {abreu2007accuracy}
\bibfield{author}{\bibinfo{person}{Rui Abreu}, \bibinfo{person}{Peter
  Zoeteweij}, {and} \bibinfo{person}{Arjan~JC Van~Gemund}.}
  \bibinfo{year}{2007}\natexlab{}.
\newblock \showarticletitle{On the accuracy of spectrum-based fault
  localization}. In \bibinfo{booktitle}{\emph{Testing: Academic and Industrial
  Conference Practice and Research Techniques-MUTATION}}. IEEE,
  \bibinfo{pages}{89--98}.
\newblock


\bibitem[\protect\citeauthoryear{Allamanis, Barr, Bird, and Sutton}{Allamanis
  et~al\mbox{.}}{2015}]%
        {allamanis2015suggesting}
\bibfield{author}{\bibinfo{person}{Miltiadis Allamanis},
  \bibinfo{person}{Earl~T Barr}, \bibinfo{person}{Christian Bird}, {and}
  \bibinfo{person}{Charles Sutton}.} \bibinfo{year}{2015}\natexlab{}.
\newblock \showarticletitle{Suggesting accurate method and class names}. In
  \bibinfo{booktitle}{\emph{Proceedings of the 10th Joint Meeting on
  Foundations of Software Engineering}}. \bibinfo{publisher}{{ACM}},
  \bibinfo{pages}{38--49}.
\newblock


\bibitem[\protect\citeauthoryear{Allamanis, Brockschmidt, and
  Khademi}{Allamanis et~al\mbox{.}}{2018}]%
        {allamanis2018learning}
\bibfield{author}{\bibinfo{person}{Miltiadis Allamanis}, \bibinfo{person}{Marc
  Brockschmidt}, {and} \bibinfo{person}{Mahmoud Khademi}.}
  \bibinfo{year}{2018}\natexlab{}.
\newblock \showarticletitle{Learning to Represent Programs with Graphs}. In
  \bibinfo{booktitle}{\emph{Proceedings of the 6th International Conference on
  Learning Representations}}. \bibinfo{publisher}{OpenReview.net}.
\newblock


\bibitem[\protect\citeauthoryear{Alon, Brody, Levy, and Yahav}{Alon
  et~al\mbox{.}}{2018a}]%
        {alon2018code2seq}
\bibfield{author}{\bibinfo{person}{Uri Alon}, \bibinfo{person}{Shaked Brody},
  \bibinfo{person}{Omer Levy}, {and} \bibinfo{person}{Eran Yahav}.}
  \bibinfo{year}{2018}\natexlab{a}.
\newblock \showarticletitle{code2seq: Generating sequences from structured
  representations of code}.
\newblock \bibinfo{journal}{\emph{arXiv preprint arXiv:1808.01400}}
  (\bibinfo{year}{2018}).
\newblock


\bibitem[\protect\citeauthoryear{Alon, Brody, Levy, and Yahav}{Alon
  et~al\mbox{.}}{2019a}]%
        {alon2019code2seq}
\bibfield{author}{\bibinfo{person}{Uri Alon}, \bibinfo{person}{Shaked Brody},
  \bibinfo{person}{Omer Levy}, {and} \bibinfo{person}{Eran Yahav}.}
  \bibinfo{year}{2019}\natexlab{a}.
\newblock \showarticletitle{code2seq: Generating Sequences from Structured
  Representations of Code}. In \bibinfo{booktitle}{\emph{Proceedings of the 7th
  International Conference on Learning Representations}}.
  \bibinfo{publisher}{OpenReview.net}.
\newblock


\bibitem[\protect\citeauthoryear{Alon, Sadaka, Levy, and Yahav}{Alon
  et~al\mbox{.}}{2020}]%
        {alon2020structural}
\bibfield{author}{\bibinfo{person}{Uri Alon}, \bibinfo{person}{Roy Sadaka},
  \bibinfo{person}{Omer Levy}, {and} \bibinfo{person}{Eran Yahav}.}
  \bibinfo{year}{2020}\natexlab{}.
\newblock \showarticletitle{Structural language models of code}. In
  \bibinfo{booktitle}{\emph{Proceedings of 37th International Conference on
  Machine Learning}}. PMLR, \bibinfo{pages}{245--256}.
\newblock


\bibitem[\protect\citeauthoryear{Alon, Zilberstein, Levy, and Yahav}{Alon
  et~al\mbox{.}}{2018b}]%
        {alon2018general}
\bibfield{author}{\bibinfo{person}{Uri Alon}, \bibinfo{person}{Meital
  Zilberstein}, \bibinfo{person}{Omer Levy}, {and} \bibinfo{person}{Eran
  Yahav}.} \bibinfo{year}{2018}\natexlab{b}.
\newblock \showarticletitle{A general path-based representation for predicting
  program properties}. In \bibinfo{booktitle}{\emph{Proceedings of the 39th ACM
  SIGPLAN Conference on Programming Language Design and Implementation}}.
  \bibinfo{publisher}{ACM}, \bibinfo{pages}{404--419}.
\newblock
\urldef\tempurl%
\url{https://doi.org/10.1145/3192366.3192412}
\showDOI{\tempurl}


\bibitem[\protect\citeauthoryear{Alon, Zilberstein, Levy, and Yahav}{Alon
  et~al\mbox{.}}{2019b}]%
        {alon2019code2vec}
\bibfield{author}{\bibinfo{person}{Uri Alon}, \bibinfo{person}{Meital
  Zilberstein}, \bibinfo{person}{Omer Levy}, {and} \bibinfo{person}{Eran
  Yahav}.} \bibinfo{year}{2019}\natexlab{b}.
\newblock \showarticletitle{code2vec: learning distributed representations of
  code}.
\newblock \bibinfo{journal}{\emph{Proceedings of the ACM on Programming
  Languages}} \bibinfo{volume}{3}, \bibinfo{number}{{POPL}}
  (\bibinfo{year}{2019}), \bibinfo{pages}{40:1--40:29}.
\newblock
\urldef\tempurl%
\url{https://doi.org/10.1145/3290353}
\showDOI{\tempurl}


\bibitem[\protect\citeauthoryear{Arcuri and Briand}{Arcuri and Briand}{2011}]%
        {random_forest}
\bibfield{author}{\bibinfo{person}{Andrea Arcuri} {and}
  \bibinfo{person}{Lionel~C. Briand}.} \bibinfo{year}{2011}\natexlab{}.
\newblock \showarticletitle{A practical guide for using statistical tests to
  assess randomized algorithms in software engineering}. In
  \bibinfo{booktitle}{\emph{Proceedings of the 33rd International Conference on
  Software Engineering}}. \bibinfo{publisher}{{ACM}}, \bibinfo{pages}{1--10}.
\newblock


\bibitem[\protect\citeauthoryear{B.~Le, Lo, Le~Goues, and Grunske}{B.~Le
  et~al\mbox{.}}{2016}]%
        {b2016learning}
\bibfield{author}{\bibinfo{person}{Tien-Duy B.~Le}, \bibinfo{person}{David Lo},
  \bibinfo{person}{Claire Le~Goues}, {and} \bibinfo{person}{Lars Grunske}.}
  \bibinfo{year}{2016}\natexlab{}.
\newblock \showarticletitle{A learning-to-rank based fault localization
  approach using likely invariants}. In \bibinfo{booktitle}{\emph{Proceedings
  of the 25th International Symposium on Software Testing and Analysis}}.
  \bibinfo{pages}{177--188}.
\newblock


\bibitem[\protect\citeauthoryear{Bavishi, Yoshida, and Prasad}{Bavishi
  et~al\mbox{.}}{2019}]%
        {bavishi2019phoenix}
\bibfield{author}{\bibinfo{person}{Rohan Bavishi}, \bibinfo{person}{Hiroaki
  Yoshida}, {and} \bibinfo{person}{Mukul~R. Prasad}.}
  \bibinfo{year}{2019}\natexlab{}.
\newblock \showarticletitle{Phoenix: Automated Data-Driven Synthesis of Repairs
  for Static Analysis Violations}. In \bibinfo{booktitle}{\emph{Proceedings of
  the 2019 27th ACM Joint Meeting on European Software Engineering Conference
  and Symposium on the Foundations of Software Engineering}}.
  \bibinfo{publisher}{{ACM}}, \bibinfo{pages}{613–624}.
\newblock


\bibitem[\protect\citeauthoryear{Benton, Li, Lou, and Zhang}{Benton
  et~al\mbox{.}}{2020}]%
        {benton2020effectiveness}
\bibfield{author}{\bibinfo{person}{Samuel Benton}, \bibinfo{person}{Xia Li},
  \bibinfo{person}{Yiling Lou}, {and} \bibinfo{person}{Lingming Zhang}.}
  \bibinfo{year}{2020}\natexlab{}.
\newblock \showarticletitle{On the Effectiveness of Unified Debugging: An
  Extensive Study on 16 Program Repair Systems}. In
  \bibinfo{booktitle}{\emph{Proceedings of the 35th IEEE/ACM International
  Conference on Automated Software Engineering}}. IEEE,
  \bibinfo{pages}{907--918}.
\newblock


\bibitem[\protect\citeauthoryear{Bhatia, Kohli, and Singh}{Bhatia
  et~al\mbox{.}}{2018}]%
        {bhatia2018neurosymbolic}
\bibfield{author}{\bibinfo{person}{Sahil Bhatia}, \bibinfo{person}{Pushmeet
  Kohli}, {and} \bibinfo{person}{Rishabh Singh}.}
  \bibinfo{year}{2018}\natexlab{}.
\newblock \showarticletitle{Neuro-Symbolic Program Corrector for Introductory
  Programming Assignments}. In \bibinfo{booktitle}{\emph{Proceedings of the
  IEEE/ACM 40th International Conference on Software Engineering}}.
  \bibinfo{pages}{60--70}.
\newblock


\bibitem[\protect\citeauthoryear{B\"{o}hme, Soremekun, Chattopadhyay,
  Ugherughe, and Zeller}{B\"{o}hme et~al\mbox{.}}{2017}]%
        {bohme2017where}
\bibfield{author}{\bibinfo{person}{Marcel B\"{o}hme},
  \bibinfo{person}{Ezekiel~O. Soremekun}, \bibinfo{person}{Sudipta
  Chattopadhyay}, \bibinfo{person}{Emamurho Ugherughe}, {and}
  \bibinfo{person}{Andreas Zeller}.} \bibinfo{year}{2017}\natexlab{}.
\newblock \showarticletitle{Where is the Bug and How is It Fixed? An Experiment
  with Practitioners}. In \bibinfo{booktitle}{\emph{Proceedings of the 11th
  Joint Meeting on Foundations of Software Engineering}}.
  \bibinfo{publisher}{{ACM}}, \bibinfo{pages}{117–128}.
\newblock


\bibitem[\protect\citeauthoryear{Briand, Labiche, and Liu}{Briand
  et~al\mbox{.}}{2007}]%
        {briand2007using}
\bibfield{author}{\bibinfo{person}{Lionel~C Briand}, \bibinfo{person}{Yvan
  Labiche}, {and} \bibinfo{person}{Xuetao Liu}.}
  \bibinfo{year}{2007}\natexlab{}.
\newblock \showarticletitle{Using machine learning to support debugging with
  tarantula}. In \bibinfo{booktitle}{\emph{Proceedings of the 18th IEEE
  International Symposium on Software Reliability}}. IEEE,
  \bibinfo{pages}{137--146}.
\newblock


\bibitem[\protect\citeauthoryear{Brockschmidt, Allamanis, Gaunt, and
  Polozov}{Brockschmidt et~al\mbox{.}}{2018}]%
        {brockschmidt2018generative}
\bibfield{author}{\bibinfo{person}{Marc Brockschmidt},
  \bibinfo{person}{Miltiadis Allamanis}, \bibinfo{person}{Alexander~L Gaunt},
  {and} \bibinfo{person}{Oleksandr Polozov}.} \bibinfo{year}{2018}\natexlab{}.
\newblock \showarticletitle{Generative code modeling with graphs}.
\newblock \bibinfo{journal}{\emph{arXiv preprint arXiv:1805.08490}}
  (\bibinfo{year}{2018}).
\newblock


\bibitem[\protect\citeauthoryear{Brody, Alon, and Yahav}{Brody
  et~al\mbox{.}}{2020}]%
        {brody2020structural}
\bibfield{author}{\bibinfo{person}{Shaked Brody}, \bibinfo{person}{Uri Alon},
  {and} \bibinfo{person}{Eran Yahav}.} \bibinfo{year}{2020}\natexlab{}.
\newblock \showarticletitle{A structural model for contextual code changes}.
\newblock \bibinfo{journal}{\emph{Proceedings of the ACM on Programming
  Languages}} \bibinfo{volume}{4}, \bibinfo{number}{OOPSLA}
  (\bibinfo{year}{2020}), \bibinfo{pages}{1--28}.
\newblock


\bibitem[\protect\citeauthoryear{Chen, Pei, and Furia}{Chen
  et~al\mbox{.}}{2017}]%
        {chen2017contract}
\bibfield{author}{\bibinfo{person}{Liushan Chen}, \bibinfo{person}{Yu Pei},
  {and} \bibinfo{person}{Carlo~A Furia}.} \bibinfo{year}{2017}\natexlab{}.
\newblock \showarticletitle{Contract-based program repair without the
  contracts}. In \bibinfo{booktitle}{\emph{Proceedings of the 32nd IEEE/ACM
  International Conference on Automated Software Engineering}}.
  \bibinfo{pages}{637--647}.
\newblock


\bibitem[\protect\citeauthoryear{Chen, Kommrusch, Tufano, Pouchet, Poshyvanyk,
  and Monperrus}{Chen et~al\mbox{.}}{2019}]%
        {chen2019sequencer}
\bibfield{author}{\bibinfo{person}{Zimin Chen}, \bibinfo{person}{Steve~James
  Kommrusch}, \bibinfo{person}{Michele Tufano}, \bibinfo{person}{Louis-No{\"e}l
  Pouchet}, \bibinfo{person}{Denys Poshyvanyk}, {and} \bibinfo{person}{Martin
  Monperrus}.} \bibinfo{year}{2019}\natexlab{}.
\newblock \showarticletitle{Sequencer: Sequence-to-sequence learning for
  end-to-end program repair}.
\newblock \bibinfo{journal}{\emph{IEEE Trans. on Software Engineering}}
  (\bibinfo{year}{2019}).
\newblock


\bibitem[\protect\citeauthoryear{Compton, Frank, Patros, and Koay}{Compton
  et~al\mbox{.}}{2020}]%
        {compton2020embedding}
\bibfield{author}{\bibinfo{person}{Rhys Compton}, \bibinfo{person}{Eibe Frank},
  \bibinfo{person}{Panos Patros}, {and} \bibinfo{person}{Abigail Koay}.}
  \bibinfo{year}{2020}\natexlab{}.
\newblock \showarticletitle{Embedding Java Classes with code2vec: Improvements
  from Variable Obfuscation}. In \bibinfo{booktitle}{\emph{Proceedings of the
  17th Mining Software Repositories}}. ACM, \bibinfo{pages}{243--253}.
\newblock


\bibitem[\protect\citeauthoryear{Falleri, Morandat, Blanc, Martinez, and
  Monperrus}{Falleri et~al\mbox{.}}{2014}]%
        {falleri2014fine}
\bibfield{author}{\bibinfo{person}{Jean-R{\'e}my Falleri},
  \bibinfo{person}{Flor{\'e}al Morandat}, \bibinfo{person}{Xavier Blanc},
  \bibinfo{person}{Matias Martinez}, {and} \bibinfo{person}{Martin Monperrus}.}
  \bibinfo{year}{2014}\natexlab{}.
\newblock \showarticletitle{Fine-grained and accurate source code
  differencing}. In \bibinfo{booktitle}{\emph{Proceedings of the 29th ACM/IEEE
  International Conference on Automated Software Engineering}}. ACM,
  \bibinfo{pages}{313--324}.
\newblock
\urldef\tempurl%
\url{https://doi.org/10.1145/2642937.2642982}
\showDOI{\tempurl}


\bibitem[\protect\citeauthoryear{Gao, Mechtaev, and Roychoudhury}{Gao
  et~al\mbox{.}}{2019}]%
        {gao2019crash}
\bibfield{author}{\bibinfo{person}{Xiang Gao}, \bibinfo{person}{Sergey
  Mechtaev}, {and} \bibinfo{person}{Abhik Roychoudhury}.}
  \bibinfo{year}{2019}\natexlab{}.
\newblock \showarticletitle{Crash-avoiding program repair}. In
  \bibinfo{booktitle}{\emph{Proceedings of the 28th ACM SIGSOFT International
  Symposium on Software Testing and Analysis}}. \bibinfo{pages}{8--18}.
\newblock


\bibitem[\protect\citeauthoryear{Ghanbari, Benton, and Zhang}{Ghanbari
  et~al\mbox{.}}{2019}]%
        {ghanbari2019practical}
\bibfield{author}{\bibinfo{person}{Ali Ghanbari}, \bibinfo{person}{Samuel
  Benton}, {and} \bibinfo{person}{Lingming Zhang}.}
  \bibinfo{year}{2019}\natexlab{}.
\newblock \showarticletitle{Practical program repair via bytecode mutation}. In
  \bibinfo{booktitle}{\emph{Proceedings of the 28th ACM SIGSOFT International
  Symposium on Software Testing and Analysis}}. ACM, \bibinfo{pages}{19--30}.
\newblock


\bibitem[\protect\citeauthoryear{Gupta, Pal, Kanade, and Shevade}{Gupta
  et~al\mbox{.}}{2017}]%
        {gupta2017deepfix}
\bibfield{author}{\bibinfo{person}{Rahul Gupta}, \bibinfo{person}{Soham Pal},
  \bibinfo{person}{Aditya Kanade}, {and} \bibinfo{person}{Shirish Shevade}.}
  \bibinfo{year}{2017}\natexlab{}.
\newblock \showarticletitle{{DeepFix:} Fixing common {C} language errors by
  deep learning}. In \bibinfo{booktitle}{\emph{Proceedings of the 31st AAAI
  Conference on Artificial Intelligence}}. \bibinfo{publisher}{{AAAI}},
  \bibinfo{pages}{1345--1351}.
\newblock


\bibitem[\protect\citeauthoryear{Hellendoorn, Sutton, Singh, Maniatis, and
  Bieber}{Hellendoorn et~al\mbox{.}}{2020}]%
        {hellendoorn2020global}
\bibfield{author}{\bibinfo{person}{Vincent~J. Hellendoorn},
  \bibinfo{person}{Charles Sutton}, \bibinfo{person}{Rishabh Singh},
  \bibinfo{person}{Petros Maniatis}, {and} \bibinfo{person}{David Bieber}.}
  \bibinfo{year}{2020}\natexlab{}.
\newblock \showarticletitle{Global Relational Models of Source Code}. In
  \bibinfo{booktitle}{\emph{Proceedings of the 8th International Conference on
  Learning Representations}}. \bibinfo{publisher}{OpenReview.net}.
\newblock


\bibitem[\protect\citeauthoryear{Hindle, Barr, Su, Gabel, and Devanbu}{Hindle
  et~al\mbox{.}}{2012}]%
        {hindle2012naturalness}
\bibfield{author}{\bibinfo{person}{Abram Hindle}, \bibinfo{person}{Earl~T.
  Barr}, \bibinfo{person}{Zhendong Su}, \bibinfo{person}{Mark Gabel}, {and}
  \bibinfo{person}{Premkumar~T. Devanbu}.} \bibinfo{year}{2012}\natexlab{}.
\newblock \showarticletitle{On the naturalness of software}. In
  \bibinfo{booktitle}{\emph{Proceedings of the 34th International Conference on
  Software Engineering}}. IEEE, \bibinfo{pages}{837--847}.
\newblock
\urldef\tempurl%
\url{https://doi.org/10.1109/ICSE.2012.6227135}
\showDOI{\tempurl}


\bibitem[\protect\citeauthoryear{Jiang, Xiong, Zhang, Gao, and Chen}{Jiang
  et~al\mbox{.}}{2018}]%
        {jiang2018shaping}
\bibfield{author}{\bibinfo{person}{Jiajun Jiang}, \bibinfo{person}{Yingfei
  Xiong}, \bibinfo{person}{Hongyu Zhang}, \bibinfo{person}{Qing Gao}, {and}
  \bibinfo{person}{Xiangqun Chen}.} \bibinfo{year}{2018}\natexlab{}.
\newblock \showarticletitle{Shaping program repair space with existing patches
  and similar code}. In \bibinfo{booktitle}{\emph{Proceedings of the 27th ACM
  SIGSOFT International Symposium on Software Testing and Analysis}}. ACM,
  \bibinfo{pages}{298--309}.
\newblock
\urldef\tempurl%
\url{https://doi.org/10.1145/3213846.3213871}
\showDOI{\tempurl}


\bibitem[\protect\citeauthoryear{Jiang, Lutellier, and Tan}{Jiang
  et~al\mbox{.}}{2021}]%
        {jiang2021cure}
\bibfield{author}{\bibinfo{person}{Nan Jiang}, \bibinfo{person}{Thibaud
  Lutellier}, {and} \bibinfo{person}{Lin Tan}.}
  \bibinfo{year}{2021}\natexlab{}.
\newblock \showarticletitle{{CURE}: Code-Aware Neural Machine Translation for
  Automatic Program Repair}. In \bibinfo{booktitle}{\emph{Proceedings of the
  43rd International Conference on Software Engineering}}.
  \bibinfo{publisher}{{IEEE}}, \bibinfo{pages}{1161--1173}.
\newblock
\urldef\tempurl%
\url{https://doi.org/10.1109/ICSE43902.2021.00107}
\showDOI{\tempurl}


\bibitem[\protect\citeauthoryear{Jones, Harrold, and Stasko}{Jones
  et~al\mbox{.}}{2002}]%
        {jones2002visualization}
\bibfield{author}{\bibinfo{person}{James~A Jones}, \bibinfo{person}{Mary~Jean
  Harrold}, {and} \bibinfo{person}{John Stasko}.}
  \bibinfo{year}{2002}\natexlab{}.
\newblock \showarticletitle{Visualization of test information to assist fault
  localization}. In \bibinfo{booktitle}{\emph{Proceedings of the 24th
  International Conference on Software Engineering}}.
  \bibinfo{pages}{467--477}.
\newblock


\bibitem[\protect\citeauthoryear{Just, Jalali, and Ernst}{Just
  et~al\mbox{.}}{2014}]%
        {just2014defects4j}
\bibfield{author}{\bibinfo{person}{Ren{\'e} Just}, \bibinfo{person}{Darioush
  Jalali}, {and} \bibinfo{person}{Michael~D Ernst}.}
  \bibinfo{year}{2014}\natexlab{}.
\newblock \showarticletitle{{Defects4J}: A database of existing faults to
  enable controlled testing studies for Java programs}. In
  \bibinfo{booktitle}{\emph{Proceedings of the 23rd International Symposium on
  Software Testing and Analysis}}. ACM, \bibinfo{pages}{437--440}.
\newblock
\urldef\tempurl%
\url{https://doi.org/10.1145/2610384.2628055}
\showDOI{\tempurl}


\bibitem[\protect\citeauthoryear{Karampatsis and Sutton}{Karampatsis and
  Sutton}{2020}]%
        {karampatsis2020how}
\bibfield{author}{\bibinfo{person}{Rafael{-}Michael Karampatsis} {and}
  \bibinfo{person}{Charles~A. Sutton}.} \bibinfo{year}{2020}\natexlab{}.
\newblock \showarticletitle{How Often Do Single-Statement Bugs Occur? The
  ManySStuBs4J Dataset}. In \bibinfo{booktitle}{\emph{Proceedings of the 17th
  Mining Software Repositories}}. \bibinfo{publisher}{{ACM}},
  \bibinfo{pages}{573--577}.
\newblock
\urldef\tempurl%
\url{https://doi.org/10.1145/3379597.3387491}
\showDOI{\tempurl}


\bibitem[\protect\citeauthoryear{Kim, Nam, Song, and Kim}{Kim
  et~al\mbox{.}}{2013}]%
        {kim2013automatic}
\bibfield{author}{\bibinfo{person}{Dongsun Kim}, \bibinfo{person}{Jaechang
  Nam}, \bibinfo{person}{Jaewoo Song}, {and} \bibinfo{person}{Sunghun Kim}.}
  \bibinfo{year}{2013}\natexlab{}.
\newblock \showarticletitle{Automatic patch generation learned from
  human-written patches}. In \bibinfo{booktitle}{\emph{Proceedings of the 35th
  International Conference on Software Engineering}}. IEEE,
  \bibinfo{pages}{802--811}.
\newblock
\urldef\tempurl%
\url{https://doi.org/10.1109/ICSE.2013.6606626}
\showDOI{\tempurl}


\bibitem[\protect\citeauthoryear{Kingma and Ba}{Kingma and Ba}{2014}]%
        {kingma2014adam}
\bibfield{author}{\bibinfo{person}{Diederik~P Kingma} {and}
  \bibinfo{person}{Jimmy Ba}.} \bibinfo{year}{2014}\natexlab{}.
\newblock \showarticletitle{Adam: A method for stochastic optimization}.
\newblock \bibinfo{journal}{\emph{arXiv preprint arXiv:1412.6980}}
  (\bibinfo{year}{2014}).
\newblock


\bibitem[\protect\citeauthoryear{Kochhar, Xia, Lo, and Li}{Kochhar
  et~al\mbox{.}}{2016}]%
        {kochhar2016practitioners}
\bibfield{author}{\bibinfo{person}{Pavneet~Singh Kochhar}, \bibinfo{person}{Xin
  Xia}, \bibinfo{person}{David Lo}, {and} \bibinfo{person}{Shanping Li}.}
  \bibinfo{year}{2016}\natexlab{}.
\newblock \showarticletitle{Practitioners' expectations on automated fault
  localization}. In \bibinfo{booktitle}{\emph{Proceedings of the 25th
  International Symposium on Software Testing and Analysis}}.
  \bibinfo{publisher}{ACM}, \bibinfo{pages}{165--176}.
\newblock


\bibitem[\protect\citeauthoryear{Kohavi}{Kohavi}{1995}]%
        {10_fold}
\bibfield{author}{\bibinfo{person}{Ron Kohavi}.}
  \bibinfo{year}{1995}\natexlab{}.
\newblock \showarticletitle{A study of cross-validation and bootstrap for
  accuracy estimation and model selection}. In
  \bibinfo{booktitle}{\emph{Proceedings of the 40th International Joint
  Conference on Artificial Intelligence}}. Montreal, Canada,
  \bibinfo{pages}{1137--1145}.
\newblock


\bibitem[\protect\citeauthoryear{Koyuncu, Bissyand{\'e}, Kim, Liu, Klein,
  Monperrus, and Traon}{Koyuncu et~al\mbox{.}}{2019}]%
        {koyuncu2019d}
\bibfield{author}{\bibinfo{person}{Anil Koyuncu},
  \bibinfo{person}{Tegawend{\'e}~F Bissyand{\'e}}, \bibinfo{person}{Dongsun
  Kim}, \bibinfo{person}{Kui Liu}, \bibinfo{person}{Jacques Klein},
  \bibinfo{person}{Martin Monperrus}, {and} \bibinfo{person}{Yves~Le Traon}.}
  \bibinfo{year}{2019}\natexlab{}.
\newblock \showarticletitle{D\&C: A Divide-and-Conquer Approach to IR-based Bug
  Localization}.
\newblock \bibinfo{journal}{\emph{arXiv preprint arXiv:1902.02703}}
  (\bibinfo{year}{2019}).
\newblock


\bibitem[\protect\citeauthoryear{Koyuncu, Liu, Bissyand{\'e}, Kim, Klein,
  Monperrus, and Traon}{Koyuncu et~al\mbox{.}}{2020}]%
        {koyuncu2020fixminer}
\bibfield{author}{\bibinfo{person}{Anil Koyuncu}, \bibinfo{person}{Kui Liu},
  \bibinfo{person}{Tegawend{\'e}~F. Bissyand{\'e}}, \bibinfo{person}{Dongsun
  Kim}, \bibinfo{person}{Jacques Klein}, \bibinfo{person}{Martin Monperrus},
  {and} \bibinfo{person}{Yves~Le Traon}.} \bibinfo{year}{2020}\natexlab{}.
\newblock \showarticletitle{{FixMiner:} Mining relevant fix patterns for
  automated program repair}.
\newblock \bibinfo{journal}{\emph{Empirical Software Engineering}}
  \bibinfo{volume}{25}, \bibinfo{number}{3} (\bibinfo{year}{2020}),
  \bibinfo{pages}{1980--2024}.
\newblock
\urldef\tempurl%
\url{https://doi.org/10.1007/s10664-019-09780-z}
\showDOI{\tempurl}


\bibitem[\protect\citeauthoryear{Le, Lo, and Le~Goues}{Le
  et~al\mbox{.}}{2016}]%
        {le2016history}
\bibfield{author}{\bibinfo{person}{Xuan Bach~D Le}, \bibinfo{person}{David Lo},
  {and} \bibinfo{person}{Claire Le~Goues}.} \bibinfo{year}{2016}\natexlab{}.
\newblock \showarticletitle{History driven program repair}. In
  \bibinfo{booktitle}{\emph{Proceedings of the 23rd IEEE International
  Conference on Software Analysis, Evolution, and Reengineering}}.
  \bibinfo{pages}{213--224}.
\newblock
\urldef\tempurl%
\url{https://doi.org/10.1109/SANER.2016.76}
\showDOI{\tempurl}


\bibitem[\protect\citeauthoryear{Le~Goues, Dewey-Vogt, Forrest, and
  Weimer}{Le~Goues et~al\mbox{.}}{2012a}]%
        {le2012systematic}
\bibfield{author}{\bibinfo{person}{Claire Le~Goues}, \bibinfo{person}{Michael
  Dewey-Vogt}, \bibinfo{person}{Stephanie Forrest}, {and}
  \bibinfo{person}{Westley Weimer}.} \bibinfo{year}{2012}\natexlab{a}.
\newblock \showarticletitle{A systematic study of automated program repair:
  Fixing 55 out of 105 bugs for \$8 each}. In
  \bibinfo{booktitle}{\emph{Proceedings of the 34th International Conference on
  Software Engineering}}. IEEE, \bibinfo{pages}{3--13}.
\newblock
\urldef\tempurl%
\url{https://doi.org/10.1109/ICSE.2012.6227211}
\showDOI{\tempurl}


\bibitem[\protect\citeauthoryear{Le~Goues, Forrest, and Weimer}{Le~Goues
  et~al\mbox{.}}{2013}]%
        {le2013current}
\bibfield{author}{\bibinfo{person}{Claire Le~Goues}, \bibinfo{person}{Stephanie
  Forrest}, {and} \bibinfo{person}{Westley Weimer}.}
  \bibinfo{year}{2013}\natexlab{}.
\newblock \showarticletitle{Current challenges in automatic software repair}.
\newblock \bibinfo{journal}{\emph{Software Quality Journal}}
  \bibinfo{volume}{21}, \bibinfo{number}{3} (\bibinfo{year}{2013}),
  \bibinfo{pages}{421--443}.
\newblock
\urldef\tempurl%
\url{https://doi.org/10.1007/s11219-013-9208-0}
\showDOI{\tempurl}


\bibitem[\protect\citeauthoryear{Le~Goues, Nguyen, Forrest, and
  Weimer}{Le~Goues et~al\mbox{.}}{2012b}]%
        {le2012genprog}
\bibfield{author}{\bibinfo{person}{Claire Le~Goues}, \bibinfo{person}{ThanhVu
  Nguyen}, \bibinfo{person}{Stephanie Forrest}, {and} \bibinfo{person}{Westley
  Weimer}.} \bibinfo{year}{2012}\natexlab{b}.
\newblock \showarticletitle{{GenProg}: A generic method for automatic software
  repair}.
\newblock \bibinfo{journal}{\emph{IEEE Transactions on Software Engineering}}
  \bibinfo{volume}{38}, \bibinfo{number}{1} (\bibinfo{year}{2012}),
  \bibinfo{pages}{54--72}.
\newblock
\urldef\tempurl%
\url{https://doi.org/10.1109/TSE.2011.104}
\showDOI{\tempurl}


\bibitem[\protect\citeauthoryear{Le~Goues, Pradel, and Roychoudhury}{Le~Goues
  et~al\mbox{.}}{2019}]%
        {le2019automated}
\bibfield{author}{\bibinfo{person}{Claire Le~Goues}, \bibinfo{person}{Michael
  Pradel}, {and} \bibinfo{person}{Abhik Roychoudhury}.}
  \bibinfo{year}{2019}\natexlab{}.
\newblock \showarticletitle{Automated Program Repair}.
\newblock \bibinfo{journal}{\emph{Commun. ACM}} \bibinfo{volume}{62},
  \bibinfo{number}{12} (\bibinfo{year}{2019}), \bibinfo{pages}{56--65}.
\newblock
\urldef\tempurl%
\url{https://doi.org/10.1145/3318162}
\showDOI{\tempurl}


\bibitem[\protect\citeauthoryear{Li, Sun, Han, and Li}{Li
  et~al\mbox{.}}{2020a}]%
        {li2020survey}
\bibfield{author}{\bibinfo{person}{Jing Li}, \bibinfo{person}{Aixin Sun},
  \bibinfo{person}{Jianglei Han}, {and} \bibinfo{person}{Chenliang Li}.}
  \bibinfo{year}{2020}\natexlab{a}.
\newblock \showarticletitle{A survey on deep learning for named entity
  recognition}.
\newblock \bibinfo{journal}{\emph{IEEE Transactions on Knowledge and Data
  Engineering}} (\bibinfo{year}{2020}).
\newblock
\urldef\tempurl%
\url{https://doi.org/10.1109/TKDE.2020.2981314}
\showDOI{\tempurl}


\bibitem[\protect\citeauthoryear{Li, Li, Zhang, and Zhang}{Li
  et~al\mbox{.}}{2019a}]%
        {li2019deepfl}
\bibfield{author}{\bibinfo{person}{Xia Li}, \bibinfo{person}{Wei Li},
  \bibinfo{person}{Yuqun Zhang}, {and} \bibinfo{person}{Lingming Zhang}.}
  \bibinfo{year}{2019}\natexlab{a}.
\newblock \showarticletitle{Deepfl: Integrating multiple fault diagnosis
  dimensions for deep fault localization}. In
  \bibinfo{booktitle}{\emph{Proceedings of the 28th ACM SIGSOFT International
  Symposium on Software Testing and Analysis}}. \bibinfo{pages}{169--180}.
\newblock


\bibitem[\protect\citeauthoryear{Li and Zhang}{Li and Zhang}{2017}]%
        {li2017transforming}
\bibfield{author}{\bibinfo{person}{Xia Li} {and} \bibinfo{person}{Lingming
  Zhang}.} \bibinfo{year}{2017}\natexlab{}.
\newblock \showarticletitle{Transforming programs and tests in tandem for fault
  localization}.
\newblock \bibinfo{journal}{\emph{Proceedings of the ACM on Programming
  Languages}} \bibinfo{volume}{1}, \bibinfo{number}{OOPSLA}
  (\bibinfo{year}{2017}), \bibinfo{pages}{1--30}.
\newblock


\bibitem[\protect\citeauthoryear{Li, Wang, and Nguyen}{Li
  et~al\mbox{.}}{2020b}]%
        {li2020dlfix}
\bibfield{author}{\bibinfo{person}{Yi Li}, \bibinfo{person}{Shaohua Wang},
  {and} \bibinfo{person}{Tien~N. Nguyen}.} \bibinfo{year}{2020}\natexlab{b}.
\newblock \showarticletitle{{DLFix}: context-based code transformation learning
  for automated program repair}. In \bibinfo{booktitle}{\emph{Proceedings of
  the 42nd International Conference on Software Engineering}}.
  \bibinfo{pages}{602--614}.
\newblock
\urldef\tempurl%
\url{https://doi.org/10.1145/3377811.3380345}
\showDOI{\tempurl}


\bibitem[\protect\citeauthoryear{Li, Wang, and Nguyen}{Li
  et~al\mbox{.}}{2021}]%
        {li2021fault}
\bibfield{author}{\bibinfo{person}{Yi Li}, \bibinfo{person}{Shaohua Wang},
  {and} \bibinfo{person}{Tien~N. Nguyen}.} \bibinfo{year}{2021}\natexlab{}.
\newblock \showarticletitle{Fault Localization with Code Coverage
  Representation Learning}. In \bibinfo{booktitle}{\emph{Proceedings of the
  43rd International Conference on Software Engineering}}.
  \bibinfo{publisher}{{IEEE}}, \bibinfo{pages}{661--673}.
\newblock


\bibitem[\protect\citeauthoryear{Li, Wang, Nguyen, and Van~Nguyen}{Li
  et~al\mbox{.}}{2019b}]%
        {li2019improving}
\bibfield{author}{\bibinfo{person}{Yi Li}, \bibinfo{person}{Shaohua Wang},
  \bibinfo{person}{Tien~N Nguyen}, {and} \bibinfo{person}{Son Van~Nguyen}.}
  \bibinfo{year}{2019}\natexlab{b}.
\newblock \showarticletitle{Improving bug detection via context-based code
  representation learning and attention-based neural networks}.
\newblock \bibinfo{journal}{\emph{Proceedings of the ACM on Programming
  Languages}} \bibinfo{volume}{3}, \bibinfo{number}{{OOPSLA}}
  (\bibinfo{year}{2019}), \bibinfo{pages}{162:1--162:30}.
\newblock
\urldef\tempurl%
\url{https://doi.org/10.1145/3360588}
\showDOI{\tempurl}


\bibitem[\protect\citeauthoryear{Liblit, Naik, Zheng, Aiken, and Jordan}{Liblit
  et~al\mbox{.}}{2005}]%
        {liblit2005scalable}
\bibfield{author}{\bibinfo{person}{Ben Liblit}, \bibinfo{person}{Mayur Naik},
  \bibinfo{person}{Alice~X Zheng}, \bibinfo{person}{Alex Aiken}, {and}
  \bibinfo{person}{Michael~I Jordan}.} \bibinfo{year}{2005}\natexlab{}.
\newblock \showarticletitle{Scalable statistical bug isolation}.
\newblock \bibinfo{journal}{\emph{ACM Sigplan Notices}} \bibinfo{volume}{40},
  \bibinfo{number}{6} (\bibinfo{year}{2005}), \bibinfo{pages}{15--26}.
\newblock


\bibitem[\protect\citeauthoryear{Lin, Koppel, Chen, and Solar-Lezama}{Lin
  et~al\mbox{.}}{2017}]%
        {lin2017quixbugs}
\bibfield{author}{\bibinfo{person}{Derrick Lin}, \bibinfo{person}{James
  Koppel}, \bibinfo{person}{Angela Chen}, {and} \bibinfo{person}{Armando
  Solar-Lezama}.} \bibinfo{year}{2017}\natexlab{}.
\newblock \showarticletitle{{QuixBugs:} A multi-lingual program repair
  benchmark set based on the Quixey Challenge}. In
  \bibinfo{booktitle}{\emph{Proceedings Companion of the 2017 ACM SIGPLAN
  International Conference on Systems, Programming, Languages, and
  Applications: Software for Humanity}}. ACM, \bibinfo{pages}{55--56}.
\newblock
\urldef\tempurl%
\url{https://doi.org/10.1145/3135932.3135941}
\showDOI{\tempurl}


\bibitem[\protect\citeauthoryear{Liu, Kim, Koyuncu, Li, Bissyand{\'e}, and
  Le~Traon}{Liu et~al\mbox{.}}{2018}]%
        {liu2018closer}
\bibfield{author}{\bibinfo{person}{Kui Liu}, \bibinfo{person}{Dongsun Kim},
  \bibinfo{person}{Anil Koyuncu}, \bibinfo{person}{Li Li},
  \bibinfo{person}{Tegawend{\'e}~F Bissyand{\'e}}, {and} \bibinfo{person}{Yves
  Le~Traon}.} \bibinfo{year}{2018}\natexlab{}.
\newblock \showarticletitle{A closer look at real-world patches}. In
  \bibinfo{booktitle}{\emph{Proceedings of the 34th International Conference on
  Software Maintenance and Evolution}}. IEEE, \bibinfo{pages}{275--286}.
\newblock
\urldef\tempurl%
\url{https://doi.org/10.1109/ICSME.2018.00037}
\showDOI{\tempurl}


\bibitem[\protect\citeauthoryear{Liu, Koyuncu, Bissyand{\'e}, Kim, Klein, and
  Traon}{Liu et~al\mbox{.}}{2019a}]%
        {liu2019you}
\bibfield{author}{\bibinfo{person}{Kui Liu}, \bibinfo{person}{Anil Koyuncu},
  \bibinfo{person}{Tegawend{\'e}~F Bissyand{\'e}}, \bibinfo{person}{Dongsun
  Kim}, \bibinfo{person}{Jacques Klein}, {and} \bibinfo{person}{Yves~Le
  Traon}.} \bibinfo{year}{2019}\natexlab{a}.
\newblock \showarticletitle{You cannot fix what you cannot find! an
  investigation of fault localization bias in benchmarking automated program
  repair systems}. In \bibinfo{booktitle}{\emph{Proceedings of the 12th IEEE
  International Conference on Software Testing, Verification and Validation}}.
  IEEE, \bibinfo{pages}{102--113}.
\newblock
\urldef\tempurl%
\url{https://doi.org/10.1109/ICST.2019.00020}
\showDOI{\tempurl}


\bibitem[\protect\citeauthoryear{Liu, Koyuncu, Kim, and Bissyand{\'e}}{Liu
  et~al\mbox{.}}{2019b}]%
        {liu2019avatar}
\bibfield{author}{\bibinfo{person}{Kui Liu}, \bibinfo{person}{Anil Koyuncu},
  \bibinfo{person}{Dongsun Kim}, {and} \bibinfo{person}{Tegawend{\'e}~F
  Bissyand{\'e}}.} \bibinfo{year}{2019}\natexlab{b}.
\newblock \showarticletitle{{AVATAR:} Fixing semantic bugs with fix patterns of
  static analysis violations}. In \bibinfo{booktitle}{\emph{Proceedings of the
  26th IEEE International Conference on Software Analysis, Evolution and
  Reengineering}}. IEEE, \bibinfo{pages}{456--467}.
\newblock
\urldef\tempurl%
\url{https://doi.org/10.1109/SANER.2019.8667970}
\showDOI{\tempurl}


\bibitem[\protect\citeauthoryear{Liu, Koyuncu, Kim, and Bissyand{\'e}}{Liu
  et~al\mbox{.}}{2019c}]%
        {liu2019tbar}
\bibfield{author}{\bibinfo{person}{Kui Liu}, \bibinfo{person}{Anil Koyuncu},
  \bibinfo{person}{Dongsun Kim}, {and} \bibinfo{person}{Tegawend{\'e}~F.
  Bissyand{\'e}}.} \bibinfo{year}{2019}\natexlab{c}.
\newblock \showarticletitle{{TBar}: Revisiting Template-based Automated Program
  Repair}. In \bibinfo{booktitle}{\emph{Proceedings of the 28th ACM SIGSOFT
  International Symposium on Software Testing and Analysis}}. ACM,
  \bibinfo{pages}{31--42}.
\newblock
\urldef\tempurl%
\url{https://doi.org/10.1145/3293882.3330577}
\showDOI{\tempurl}


\bibitem[\protect\citeauthoryear{Liu, Wang, Koyuncu, Kim, Bissyand{\'e}, Kim,
  Wu, Klein, Mao, and Traon}{Liu et~al\mbox{.}}{2020}]%
        {liu2020efficiency}
\bibfield{author}{\bibinfo{person}{Kui Liu}, \bibinfo{person}{Shangwen Wang},
  \bibinfo{person}{Anil Koyuncu}, \bibinfo{person}{Kisub Kim},
  \bibinfo{person}{Tegawend{\'e}~F. Bissyand{\'e}}, \bibinfo{person}{Dongsun
  Kim}, \bibinfo{person}{Peng Wu}, \bibinfo{person}{Jacques Klein},
  \bibinfo{person}{Xiaoguang Mao}, {and} \bibinfo{person}{Yves~Le Traon}.}
  \bibinfo{year}{2020}\natexlab{}.
\newblock \showarticletitle{On the Efficiency of Test Suite based Program
  Repair: A Systematic Assessment of 16 Automated Repair Systems for Java
  Programs}. In \bibinfo{booktitle}{\emph{Proceedings of the 42nd International
  Conference on Software Engineering}}. ACM, \bibinfo{pages}{615--627}.
\newblock
\urldef\tempurl%
\url{https://doi.org/10.1145/3377811.3380338}
\showDOI{\tempurl}


\bibitem[\protect\citeauthoryear{Long and Rinard}{Long and Rinard}{2015}]%
        {long2015staged}
\bibfield{author}{\bibinfo{person}{Fan Long} {and} \bibinfo{person}{Martin
  Rinard}.} \bibinfo{year}{2015}\natexlab{}.
\newblock \showarticletitle{Staged program repair with condition synthesis}. In
  \bibinfo{booktitle}{\emph{Proceedings of the 10th Joint Meeting on
  Foundations of Software Engineering}}. ACM, \bibinfo{pages}{166--178}.
\newblock
\urldef\tempurl%
\url{https://doi.org/10.1145/2786805.2786811}
\showDOI{\tempurl}


\bibitem[\protect\citeauthoryear{Long and Rinard}{Long and Rinard}{2016a}]%
        {long2016analysis}
\bibfield{author}{\bibinfo{person}{Fan Long} {and} \bibinfo{person}{Martin
  Rinard}.} \bibinfo{year}{2016}\natexlab{a}.
\newblock \showarticletitle{An analysis of the search spaces for generate and
  validate patch generation systems}. In \bibinfo{booktitle}{\emph{Proceedings
  of the 38th International Conference on Software Engineering}}. IEEE,
  \bibinfo{pages}{702--713}.
\newblock
\urldef\tempurl%
\url{https://doi.org/10.1145/2884781.2884872}
\showDOI{\tempurl}


\bibitem[\protect\citeauthoryear{Long and Rinard}{Long and Rinard}{2016b}]%
        {long2016automatic}
\bibfield{author}{\bibinfo{person}{Fan Long} {and} \bibinfo{person}{Martin
  Rinard}.} \bibinfo{year}{2016}\natexlab{b}.
\newblock \showarticletitle{Automatic patch generation by learning correct
  code}. In \bibinfo{booktitle}{\emph{Proceedings of the 43rd Annual {ACM}
  {SIGPLAN-SIGACT} Symposium on Principles of Programming Languages}},
  Vol.~\bibinfo{volume}{51}. ACM, \bibinfo{pages}{298--312}.
\newblock
\urldef\tempurl%
\url{https://doi.org/10.1145/2837614.2837617}
\showDOI{\tempurl}


\bibitem[\protect\citeauthoryear{Lou, Ghanbari, Li, Zhang, Zhang, Hao, and
  Zhang}{Lou et~al\mbox{.}}{2020}]%
        {lou2020can}
\bibfield{author}{\bibinfo{person}{Yiling Lou}, \bibinfo{person}{Ali Ghanbari},
  \bibinfo{person}{Xia Li}, \bibinfo{person}{Lingming Zhang},
  \bibinfo{person}{Haotian Zhang}, \bibinfo{person}{Dan Hao}, {and}
  \bibinfo{person}{Lu Zhang}.} \bibinfo{year}{2020}\natexlab{}.
\newblock \showarticletitle{Can automated program repair refine fault
  localization? a unified debugging approach}. In
  \bibinfo{booktitle}{\emph{Proceedings of the 29th ACM SIGSOFT International
  Symposium on Software Testing and Analysis}}. \bibinfo{publisher}{{ACM}},
  \bibinfo{pages}{75--87}.
\newblock
\urldef\tempurl%
\url{https://doi.org/10.1145/3395363.3397351}
\showDOI{\tempurl}


\bibitem[\protect\citeauthoryear{Lutellier, Pham, Pang, Li, Wei, and
  Tan}{Lutellier et~al\mbox{.}}{2020}]%
        {lutellier2020coconut}
\bibfield{author}{\bibinfo{person}{Thibaud Lutellier},
  \bibinfo{person}{Hung~Viet Pham}, \bibinfo{person}{Lawrence Pang},
  \bibinfo{person}{Yitong Li}, \bibinfo{person}{Moshi Wei}, {and}
  \bibinfo{person}{Lin Tan}.} \bibinfo{year}{2020}\natexlab{}.
\newblock \showarticletitle{{CoCoNuT}: combining context-aware neural
  translation models using ensemble for program repair}. In
  \bibinfo{booktitle}{\emph{Proceedings of the 29th ACM SIGSOFT International
  Symposium on Software Testing and Analysis}}. \bibinfo{publisher}{{ACM}},
  \bibinfo{pages}{101--114}.
\newblock
\urldef\tempurl%
\url{https://doi.org/10.1145/3395363.3397369}
\showDOI{\tempurl}


\bibitem[\protect\citeauthoryear{Madeiral, Urli, Maia, and Monperrus}{Madeiral
  et~al\mbox{.}}{2019}]%
        {madeiral2019bears}
\bibfield{author}{\bibinfo{person}{Fernanda Madeiral}, \bibinfo{person}{Simon
  Urli}, \bibinfo{person}{Marcelo Maia}, {and} \bibinfo{person}{Martin
  Monperrus}.} \bibinfo{year}{2019}\natexlab{}.
\newblock \showarticletitle{{BEARS:} An Extensible Java Bug Benchmark for
  Automatic Program Repair Studies}. In \bibinfo{booktitle}{\emph{Proceedings
  of the 26th International Conference on Software Analysis, Evolution and
  Reengineering}}. IEEE, \bibinfo{pages}{468--478}.
\newblock
\urldef\tempurl%
\url{https://doi.org/10.1109/SANER.2019.8667991}
\showDOI{\tempurl}


\bibitem[\protect\citeauthoryear{Martinez and Monperrus}{Martinez and
  Monperrus}{2016}]%
        {martinez2016astor}
\bibfield{author}{\bibinfo{person}{Matias Martinez} {and}
  \bibinfo{person}{Martin Monperrus}.} \bibinfo{year}{2016}\natexlab{}.
\newblock \showarticletitle{{ASTOR:} a program repair library for Java (demo)}.
  In \bibinfo{booktitle}{\emph{Proceedings of the 25th International Symposium
  on Software Testing and Analysis}}. ACM, \bibinfo{pages}{441--444}.
\newblock
\urldef\tempurl%
\url{https://doi.org/10.1145/2931037.2948705}
\showDOI{\tempurl}


\bibitem[\protect\citeauthoryear{Mayer and Stumptner}{Mayer and
  Stumptner}{2008}]%
        {mayer2008evaluating}
\bibfield{author}{\bibinfo{person}{Wolfgang Mayer} {and}
  \bibinfo{person}{Markus Stumptner}.} \bibinfo{year}{2008}\natexlab{}.
\newblock \showarticletitle{Evaluating models for model-based debugging}. In
  \bibinfo{booktitle}{\emph{Proceedings of the 23rd IEEE/ACM International
  Conference on Automated Software Engineering}}. IEEE,
  \bibinfo{pages}{128--137}.
\newblock


\bibitem[\protect\citeauthoryear{Mechtaev, Yi, and Roychoudhury}{Mechtaev
  et~al\mbox{.}}{2016}]%
        {mechtaev2016angelix}
\bibfield{author}{\bibinfo{person}{Sergey Mechtaev}, \bibinfo{person}{Jooyong
  Yi}, {and} \bibinfo{person}{Abhik Roychoudhury}.}
  \bibinfo{year}{2016}\natexlab{}.
\newblock \showarticletitle{Angelix: Scalable multiline program patch synthesis
  via symbolic analysis}. In \bibinfo{booktitle}{\emph{Proceedings of the 38th
  International Conference on Software Engineering}}. ACM,
  \bibinfo{pages}{691--701}.
\newblock
\urldef\tempurl%
\url{https://doi.org/10.1145/2884781.2884807}
\showDOI{\tempurl}


\bibitem[\protect\citeauthoryear{Mesbah, Rice, Johnston, Glorioso, and
  Aftandilian}{Mesbah et~al\mbox{.}}{2019}]%
        {mesbah2019deepdelta}
\bibfield{author}{\bibinfo{person}{Ali Mesbah}, \bibinfo{person}{Andrew Rice},
  \bibinfo{person}{Emily Johnston}, \bibinfo{person}{Nick Glorioso}, {and}
  \bibinfo{person}{Edward Aftandilian}.} \bibinfo{year}{2019}\natexlab{}.
\newblock \showarticletitle{{DeepDelta}: learning to repair compilation
  errors}. In \bibinfo{booktitle}{\emph{Proceedings of the 2019 27th ACM Joint
  Meeting on European Software Engineering Conference and Symposium on the
  Foundations of Software Engineering}}. \bibinfo{pages}{925--936}.
\newblock


\bibitem[\protect\citeauthoryear{Nguyen, Qi, Roychoudhury, and Chandra}{Nguyen
  et~al\mbox{.}}{2013}]%
        {nguyen2013semfix}
\bibfield{author}{\bibinfo{person}{Hoang Duong~Thien Nguyen},
  \bibinfo{person}{Dawei Qi}, \bibinfo{person}{Abhik Roychoudhury}, {and}
  \bibinfo{person}{Satish Chandra}.} \bibinfo{year}{2013}\natexlab{}.
\newblock \showarticletitle{{SemFix}: Program repair via semantic analysis}. In
  \bibinfo{booktitle}{\emph{Proceedings of the 35th International Conference on
  Software Engineering}}. IEEE, \bibinfo{pages}{772--781}.
\newblock
\urldef\tempurl%
\url{https://doi.org/10.1109/ICSE.2013.6606623}
\showDOI{\tempurl}


\bibitem[\protect\citeauthoryear{Papadakis and Le~Traon}{Papadakis and
  Le~Traon}{2015}]%
        {papadakis2015metallaxis}
\bibfield{author}{\bibinfo{person}{Mike Papadakis} {and} \bibinfo{person}{Yves
  Le~Traon}.} \bibinfo{year}{2015}\natexlab{}.
\newblock \showarticletitle{Metallaxis-FL: mutation-based fault localization}.
\newblock \bibinfo{journal}{\emph{Software Testing, Verification and
  Reliability}} \bibinfo{volume}{25}, \bibinfo{number}{5-7}
  (\bibinfo{year}{2015}), \bibinfo{pages}{605--628}.
\newblock


\bibitem[\protect\citeauthoryear{Parnin and Orso}{Parnin and Orso}{2011}]%
        {parnin2011automated}
\bibfield{author}{\bibinfo{person}{Chris Parnin} {and}
  \bibinfo{person}{Alessandro Orso}.} \bibinfo{year}{2011}\natexlab{}.
\newblock \showarticletitle{Are automated debugging techniques actually helping
  programmers?}. In \bibinfo{booktitle}{\emph{Proceedings of the 2011
  International Symposium on Software Testing and Analysis}}.
  \bibinfo{pages}{199--209}.
\newblock


\bibitem[\protect\citeauthoryear{Pearson, Campos, Just, Fraser, Abreu, Ernst,
  Pang, and Keller}{Pearson et~al\mbox{.}}{2017}]%
        {pearson2017evaluating}
\bibfield{author}{\bibinfo{person}{Spencer Pearson}, \bibinfo{person}{Jos{\'e}
  Campos}, \bibinfo{person}{Ren{\'e} Just}, \bibinfo{person}{Gordon Fraser},
  \bibinfo{person}{Rui Abreu}, \bibinfo{person}{Michael~D. Ernst},
  \bibinfo{person}{Deric Pang}, {and} \bibinfo{person}{Benjamin Keller}.}
  \bibinfo{year}{2017}\natexlab{}.
\newblock \showarticletitle{Evaluating and improving fault localization}. In
  \bibinfo{booktitle}{\emph{Proceedings of the 39th International Conference on
  Software Engineering}}. \bibinfo{publisher}{{ACM}},
  \bibinfo{pages}{609--620}.
\newblock
\urldef\tempurl%
\url{https://doi.org/10.1109/ICSE.2017.62}
\showDOI{\tempurl}


\bibitem[\protect\citeauthoryear{Qi, Long, Achour, and Rinard}{Qi
  et~al\mbox{.}}{2015}]%
        {qi2015analysis}
\bibfield{author}{\bibinfo{person}{Zichao Qi}, \bibinfo{person}{Fan Long},
  \bibinfo{person}{Sara Achour}, {and} \bibinfo{person}{Martin Rinard}.}
  \bibinfo{year}{2015}\natexlab{}.
\newblock \showarticletitle{An analysis of patch plausibility and correctness
  for generate-and-validate patch generation systems}. In
  \bibinfo{booktitle}{\emph{Proceedings of the 24th International Symposium on
  Software Testing and Analysis}}. ACM, \bibinfo{pages}{24--36}.
\newblock
\urldef\tempurl%
\url{https://doi.org/10.1145/2771783.2771791}
\showDOI{\tempurl}


\bibitem[\protect\citeauthoryear{Rubinstein}{Rubinstein}{1999}]%
        {rubinstein1999the}
\bibfield{author}{\bibinfo{person}{Reuven~Y. Rubinstein}.}
  \bibinfo{year}{1999}\natexlab{}.
\newblock \showarticletitle{The Cross-Entropy Method for Combinatorial and
  Continuous Optimization}.
\newblock \bibinfo{journal}{\emph{Methodology And Computing In Applied
  Probability}}  \bibinfo{volume}{1} (\bibinfo{year}{1999}),
  \bibinfo{pages}{127--190}.
\newblock


\bibitem[\protect\citeauthoryear{Saha, Lyu, Lam, Yoshida, and Prasad}{Saha
  et~al\mbox{.}}{2018}]%
        {saha2018bugs}
\bibfield{author}{\bibinfo{person}{Ripon Saha}, \bibinfo{person}{Yingjun Lyu},
  \bibinfo{person}{Wing Lam}, \bibinfo{person}{Hiroaki Yoshida}, {and}
  \bibinfo{person}{Mukul~R. Prasad}.} \bibinfo{year}{2018}\natexlab{}.
\newblock \showarticletitle{Bugs.jar: A large-scale, diverse dataset of
  real-world java bugs}. In \bibinfo{booktitle}{\emph{Proceedings of the 15th
  IEEE/ACM International Conference on Mining Software Repositories}}. ACM,
  \bibinfo{pages}{10--13}.
\newblock
\urldef\tempurl%
\url{https://doi.org/10.1145/3196398.3196473}
\showDOI{\tempurl}


\bibitem[\protect\citeauthoryear{Santos, Campbell, Patel, Hindle, and
  Amaral}{Santos et~al\mbox{.}}{2018}]%
        {santos2018syntax}
\bibfield{author}{\bibinfo{person}{Eddie~Antonio Santos},
  \bibinfo{person}{Hazel~Victoria Campbell}, \bibinfo{person}{Dhvani Patel},
  \bibinfo{person}{Abram Hindle}, {and} \bibinfo{person}{José~Nelson Amaral}.}
  \bibinfo{year}{2018}\natexlab{}.
\newblock \showarticletitle{Syntax and sensibility: Using language models to
  detect and correct syntax errors}. In \bibinfo{booktitle}{\emph{Proceedings
  of the IEEE 25th International Conference on Software Analysis, Evolution and
  Reengineering}}. \bibinfo{pages}{311--322}.
\newblock


\bibitem[\protect\citeauthoryear{Santos, Campbell, Hindle, and Amaral}{Santos
  et~al\mbox{.}}{2017}]%
        {santos2017finding}
\bibfield{author}{\bibinfo{person}{Eddie~A Santos}, \bibinfo{person}{Joshua~C
  Campbell}, \bibinfo{person}{Abram Hindle}, {and}
  \bibinfo{person}{Jos{\'e}~Nelson Amaral}.} \bibinfo{year}{2017}\natexlab{}.
\newblock \showarticletitle{Finding and correcting syntax errors using
  recurrent neural networks}.
\newblock \bibinfo{journal}{\emph{PeerJ Preprints}}  \bibinfo{volume}{5}
  (\bibinfo{year}{2017}), \bibinfo{pages}{e3123v1}.
\newblock
\urldef\tempurl%
\url{https://doi.org/10.7287/peerj.preprints.3123v1}
\showDOI{\tempurl}


\bibitem[\protect\citeauthoryear{Shariffdeen, Tan, Gao, and
  Roychoudhury}{Shariffdeen et~al\mbox{.}}{2021}]%
        {shariffdeen2021automated}
\bibfield{author}{\bibinfo{person}{Ridwan~Salihin Shariffdeen},
  \bibinfo{person}{Shin~Hwei Tan}, \bibinfo{person}{Mingyuan Gao}, {and}
  \bibinfo{person}{Abhik Roychoudhury}.} \bibinfo{year}{2021}\natexlab{}.
\newblock \showarticletitle{Automated Patch Transplantation}.
\newblock \bibinfo{journal}{\emph{ACM Transactions on Software Engineering and
  Methodology}} \bibinfo{volume}{30}, \bibinfo{number}{1}
  (\bibinfo{year}{2021}), \bibinfo{pages}{6:1--6:36}.
\newblock


\bibitem[\protect\citeauthoryear{Smith, Barr, Le~Goues, and Brun}{Smith
  et~al\mbox{.}}{2015}]%
        {smith2015cure}
\bibfield{author}{\bibinfo{person}{Edward~K Smith}, \bibinfo{person}{Earl~T
  Barr}, \bibinfo{person}{Claire Le~Goues}, {and} \bibinfo{person}{Yuriy
  Brun}.} \bibinfo{year}{2015}\natexlab{}.
\newblock \showarticletitle{Is the cure worse than the disease? overfitting in
  automated program repair}. In \bibinfo{booktitle}{\emph{Proceedings of the
  10th Joint Meeting on Foundations of Software Engineering}}. ACM,
  \bibinfo{pages}{532--543}.
\newblock
\urldef\tempurl%
\url{https://doi.org/10.1145/2786805.2786825}
\showDOI{\tempurl}


\bibitem[\protect\citeauthoryear{Sohn and Yoo}{Sohn and Yoo}{2017}]%
        {sohn2017fluccs}
\bibfield{author}{\bibinfo{person}{Jeongju Sohn} {and} \bibinfo{person}{Shin
  Yoo}.} \bibinfo{year}{2017}\natexlab{}.
\newblock \showarticletitle{Fluccs: Using code and change metrics to improve
  fault localization}. In \bibinfo{booktitle}{\emph{Proceedings of the 26th ACM
  SIGSOFT International Symposium on Software Testing and Analysis}}.
  \bibinfo{pages}{273--283}.
\newblock


\bibitem[\protect\citeauthoryear{Sutskever, Vinyals, and Le}{Sutskever
  et~al\mbox{.}}{2014}]%
        {sutskever2014sequence}
\bibfield{author}{\bibinfo{person}{Ilya Sutskever}, \bibinfo{person}{Oriol
  Vinyals}, {and} \bibinfo{person}{Quoc~V Le}.}
  \bibinfo{year}{2014}\natexlab{}.
\newblock \showarticletitle{Sequence to sequence learning with neural
  networks}. In \bibinfo{booktitle}{\emph{Advances in Neural Information
  Processing Systems}}. \bibinfo{pages}{3104--3112}.
\newblock


\bibitem[\protect\citeauthoryear{Tan and Roychoudhury}{Tan and
  Roychoudhury}{2015}]%
        {tan2015relifix}
\bibfield{author}{\bibinfo{person}{Shin~Hwei Tan} {and} \bibinfo{person}{Abhik
  Roychoudhury}.} \bibinfo{year}{2015}\natexlab{}.
\newblock \showarticletitle{relifix: Automated repair of software regressions}.
  In \bibinfo{booktitle}{\emph{Proceedings of the IEEE/ACM 37th IEEE
  International Conference on Software Engineering}}, Vol.~\bibinfo{volume}{1}.
  IEEE, \bibinfo{pages}{471--482}.
\newblock


\bibitem[\protect\citeauthoryear{Tan, Yoshida, Prasad, and Roychoudhury}{Tan
  et~al\mbox{.}}{2016}]%
        {tan2016anti}
\bibfield{author}{\bibinfo{person}{Shin~Hwei Tan}, \bibinfo{person}{Hiroaki
  Yoshida}, \bibinfo{person}{Mukul~R Prasad}, {and} \bibinfo{person}{Abhik
  Roychoudhury}.} \bibinfo{year}{2016}\natexlab{}.
\newblock \showarticletitle{Anti-patterns in search-based program repair}. In
  \bibinfo{booktitle}{\emph{Proceedings of the 2016 24th ACM SIGSOFT
  International Symposium on Foundations of Software Engineering}}.
  \bibinfo{publisher}{{ACM}}, \bibinfo{pages}{727--738}.
\newblock


\bibitem[\protect\citeauthoryear{Tao, Kim, Kim, and Xu}{Tao
  et~al\mbox{.}}{2014}]%
        {tao2014automatically}
\bibfield{author}{\bibinfo{person}{Yida Tao}, \bibinfo{person}{Jindae Kim},
  \bibinfo{person}{Sunghun Kim}, {and} \bibinfo{person}{Chang Xu}.}
  \bibinfo{year}{2014}\natexlab{}.
\newblock \showarticletitle{Automatically generated patches as debugging aids:
  a human study}. In \bibinfo{booktitle}{\emph{Proceedings of the 22nd ACM
  SIGSOFT International Symposium on Foundations of Software Engineering}}.
  \bibinfo{publisher}{ACM}, \bibinfo{pages}{64--74}.
\newblock


\bibitem[\protect\citeauthoryear{Tian, Liu, Kabor{\'e}, Koyuncu, Li, Klein, and
  Bissyand{\'e}}{Tian et~al\mbox{.}}{2020}]%
        {tian2020evaluating}
\bibfield{author}{\bibinfo{person}{Haoye Tian}, \bibinfo{person}{Kui Liu},
  \bibinfo{person}{Abdoul~Kader Kabor{\'e}}, \bibinfo{person}{Anil Koyuncu},
  \bibinfo{person}{Li Li}, \bibinfo{person}{Jacques Klein}, {and}
  \bibinfo{person}{Tegawend{\'e}~F. Bissyand{\'e}}.}
  \bibinfo{year}{2020}\natexlab{}.
\newblock \showarticletitle{Evaluating Representation Learning of Code Changes
  for Predicting Patch Correctness in Program Repair}. In
  \bibinfo{booktitle}{\emph{Proceedings of the 35th IEEE/ACM International
  Conference on Automated Software Engineering}}. \bibinfo{publisher}{{IEEE}},
  \bibinfo{pages}{981--992}.
\newblock


\bibitem[\protect\citeauthoryear{Tufano, Watson, Bavota, Di~Penta, White, and
  Poshyvanyk}{Tufano et~al\mbox{.}}{2019}]%
        {tufano2019empirical}
\bibfield{author}{\bibinfo{person}{Michele Tufano}, \bibinfo{person}{Cody
  Watson}, \bibinfo{person}{Gabriele Bavota}, \bibinfo{person}{Massimiliano
  Di~Penta}, \bibinfo{person}{Martin White}, {and} \bibinfo{person}{Denys
  Poshyvanyk}.} \bibinfo{year}{2019}\natexlab{}.
\newblock \showarticletitle{An empirical study on learning bug-fixing patches
  in the wild via neural machine translation}.
\newblock \bibinfo{journal}{\emph{ACM Transactions on Software Engineering and
  Methodology}} \bibinfo{volume}{28}, \bibinfo{number}{4}
  (\bibinfo{year}{2019}), \bibinfo{pages}{19:1--19:29}.
\newblock
\urldef\tempurl%
\url{https://doi.org/10.1145/3340544}
\showDOI{\tempurl}


\bibitem[\protect\citeauthoryear{Vasic, Kanade, Maniatis, Bieber, and
  Singh}{Vasic et~al\mbox{.}}{2019}]%
        {vasic2019neural}
\bibfield{author}{\bibinfo{person}{Marko Vasic}, \bibinfo{person}{Aditya
  Kanade}, \bibinfo{person}{Petros Maniatis}, \bibinfo{person}{David Bieber},
  {and} \bibinfo{person}{Rishabh Singh}.} \bibinfo{year}{2019}\natexlab{}.
\newblock \showarticletitle{Neural program repair by jointly learning to
  localize and repair}.
\newblock \bibinfo{journal}{\emph{arXiv preprint arXiv:1904.01720}}
  (\bibinfo{year}{2019}).
\newblock


\bibitem[\protect\citeauthoryear{Vinyals, Fortunato, and Jaitly}{Vinyals
  et~al\mbox{.}}{2015}]%
        {vinyals2015pointer}
\bibfield{author}{\bibinfo{person}{Oriol Vinyals}, \bibinfo{person}{Meire
  Fortunato}, {and} \bibinfo{person}{Navdeep Jaitly}.}
  \bibinfo{year}{2015}\natexlab{}.
\newblock \showarticletitle{Pointer networks}. In
  \bibinfo{booktitle}{\emph{Proceedings of the 29th Advances in Neural
  Information Processing Systems}}. \bibinfo{pages}{2692--2700}.
\newblock


\bibitem[\protect\citeauthoryear{Wang, Wen, Lin, Wu, Qin, Zou, Mao, and
  Jin}{Wang et~al\mbox{.}}{2020}]%
        {wang2020automated}
\bibfield{author}{\bibinfo{person}{Shangwen Wang}, \bibinfo{person}{Ming Wen},
  \bibinfo{person}{Bo Lin}, \bibinfo{person}{Hongjun Wu},
  \bibinfo{person}{Yihao Qin}, \bibinfo{person}{Deqing Zou},
  \bibinfo{person}{Xiaoguang Mao}, {and} \bibinfo{person}{Hai Jin}.}
  \bibinfo{year}{2020}\natexlab{}.
\newblock \showarticletitle{Automated Patch Correctness Assessment: How Far are
  We?}. In \bibinfo{booktitle}{\emph{Proceedings of the 35th IEEE/ACM
  International Conference on Automated Software Engineering}}.
  \bibinfo{publisher}{{IEEE}}, \bibinfo{pages}{968--980}.
\newblock


\bibitem[\protect\citeauthoryear{Weimer, Nguyen, Le~Goues, and Forrest}{Weimer
  et~al\mbox{.}}{2009}]%
        {weimer2009automatically}
\bibfield{author}{\bibinfo{person}{Westley Weimer}, \bibinfo{person}{ThanhVu
  Nguyen}, \bibinfo{person}{Claire Le~Goues}, {and} \bibinfo{person}{Stephanie
  Forrest}.} \bibinfo{year}{2009}\natexlab{}.
\newblock \showarticletitle{Automatically finding patches using genetic
  programming}. In \bibinfo{booktitle}{\emph{Proceedings of the 31st
  International Conference on Software Engineering}}. IEEE,
  \bibinfo{pages}{364--374}.
\newblock
\urldef\tempurl%
\url{https://doi.org/10.1109/ICSE.2009.5070536}
\showDOI{\tempurl}


\bibitem[\protect\citeauthoryear{Wen, Chen, Tian, Wu, and Cheung}{Wen
  et~al\mbox{.}}{2019}]%
        {wen2019historical}
\bibfield{author}{\bibinfo{person}{Ming Wen}, \bibinfo{person}{Junjie Chen},
  \bibinfo{person}{Yongqiang Tian}, \bibinfo{person}{Rongxin Wu}, {and}
  \bibinfo{person}{Shing-Chi Cheung}.} \bibinfo{year}{2019}\natexlab{}.
\newblock \showarticletitle{Historical Spectrum based Fault Localization}.
\newblock \bibinfo{journal}{\emph{IEEE Transactions on Software Engineering}}
  \bibinfo{volume}{PP}, \bibinfo{number}{99} (\bibinfo{year}{2019}),
  \bibinfo{pages}{1--1}.
\newblock


\bibitem[\protect\citeauthoryear{Wen, Chen, Wu, Hao, and Cheung}{Wen
  et~al\mbox{.}}{2018}]%
        {wen2018context}
\bibfield{author}{\bibinfo{person}{Ming Wen}, \bibinfo{person}{Junjie Chen},
  \bibinfo{person}{Rongxin Wu}, \bibinfo{person}{Dan Hao}, {and}
  \bibinfo{person}{Shing-Chi Cheung}.} \bibinfo{year}{2018}\natexlab{}.
\newblock \showarticletitle{Context-aware patch generation for better automated
  program repair}. In \bibinfo{booktitle}{\emph{Proceedings of the 40th
  International Conference on Software Engineering}}. ACM,
  \bibinfo{pages}{1--11}.
\newblock
\urldef\tempurl%
\url{https://doi.org/10.1145/3180155.3180233}
\showDOI{\tempurl}


\bibitem[\protect\citeauthoryear{Wong, Debroy, Golden, Xu, and
  Thuraisingham}{Wong et~al\mbox{.}}{2011}]%
        {wong2011effective}
\bibfield{author}{\bibinfo{person}{W.~Eric Wong}, \bibinfo{person}{Vidroha
  Debroy}, \bibinfo{person}{Richard Golden}, \bibinfo{person}{Xiaofeng Xu},
  {and} \bibinfo{person}{Bhavani Thuraisingham}.}
  \bibinfo{year}{2011}\natexlab{}.
\newblock \showarticletitle{Effective software fault localization using an RBF
  neural network}.
\newblock \bibinfo{journal}{\emph{IEEE Transactions on Reliability}}
  \bibinfo{volume}{61}, \bibinfo{number}{1} (\bibinfo{year}{2011}),
  \bibinfo{pages}{149--169}.
\newblock


\bibitem[\protect\citeauthoryear{Wong, Gao, Li, Abreu, and Wotawa}{Wong
  et~al\mbox{.}}{2016}]%
        {wong2016survey}
\bibfield{author}{\bibinfo{person}{W.~Eric Wong}, \bibinfo{person}{Ruizhi Gao},
  \bibinfo{person}{Yihao Li}, \bibinfo{person}{Rui Abreu}, {and}
  \bibinfo{person}{Franz Wotawa}.} \bibinfo{year}{2016}\natexlab{}.
\newblock \showarticletitle{A Survey on Software Fault Localization}.
\newblock \bibinfo{journal}{\emph{{IEEE} Transactions on Software Engineering}}
  \bibinfo{volume}{42}, \bibinfo{number}{8} (\bibinfo{year}{2016}),
  \bibinfo{pages}{707--740}.
\newblock
\urldef\tempurl%
\url{https://doi.org/10.1109/TSE.2016.2521368}
\showDOI{\tempurl}


\bibitem[\protect\citeauthoryear{Wong and Qi}{Wong and Qi}{2009}]%
        {wong2009bp}
\bibfield{author}{\bibinfo{person}{W.~Eric Wong} {and} \bibinfo{person}{Yu
  Qi}.} \bibinfo{year}{2009}\natexlab{}.
\newblock \showarticletitle{BP neural network-based effective fault
  localization}.
\newblock \bibinfo{journal}{\emph{International Journal of Software Engineering
  and Knowledge Engineering}} \bibinfo{volume}{19}, \bibinfo{number}{04}
  (\bibinfo{year}{2009}), \bibinfo{pages}{573--597}.
\newblock


\bibitem[\protect\citeauthoryear{Xie, Liu, Song, Chen, Xuan, and Xu}{Xie
  et~al\mbox{.}}{2016}]%
        {xie2016revisit}
\bibfield{author}{\bibinfo{person}{Xiaoyuan Xie}, \bibinfo{person}{Zicong Liu},
  \bibinfo{person}{Shuo Song}, \bibinfo{person}{Zhenyu Chen},
  \bibinfo{person}{Jifeng Xuan}, {and} \bibinfo{person}{Baowen Xu}.}
  \bibinfo{year}{2016}\natexlab{}.
\newblock \showarticletitle{Revisit of automatic debugging via human
  focus-tracking analysis}. In \bibinfo{booktitle}{\emph{Proceedings of the
  38th International Conference on Software Engineering}}.
  \bibinfo{publisher}{{ACM}}, \bibinfo{pages}{808--819}.
\newblock


\bibitem[\protect\citeauthoryear{Xin and Reiss}{Xin and Reiss}{2017}]%
        {xin2017leveraging}
\bibfield{author}{\bibinfo{person}{Qi Xin} {and} \bibinfo{person}{Steven~P
  Reiss}.} \bibinfo{year}{2017}\natexlab{}.
\newblock \showarticletitle{Leveraging syntax-related code for automated
  program repair}. In \bibinfo{booktitle}{\emph{Proceedings of the 32nd
  IEEE/ACM International Conference on Automated Software Engineering}}.
  \bibinfo{pages}{660--670}.
\newblock
\urldef\tempurl%
\url{https://doi.org/10.1109/ASE.2017.8115676}
\showDOI{\tempurl}


\bibitem[\protect\citeauthoryear{Xiong, Liu, Zeng, Zhang, and Huang}{Xiong
  et~al\mbox{.}}{2018}]%
        {xiong2018identifying}
\bibfield{author}{\bibinfo{person}{Yingfei Xiong}, \bibinfo{person}{Xinyuan
  Liu}, \bibinfo{person}{Muhan Zeng}, \bibinfo{person}{Lu Zhang}, {and}
  \bibinfo{person}{Gang Huang}.} \bibinfo{year}{2018}\natexlab{}.
\newblock \showarticletitle{Identifying patch correctness in test-based program
  repair}. In \bibinfo{booktitle}{\emph{Proceedings of the 40th International
  Conference on Software Engineering}}. ACM, \bibinfo{pages}{789--799}.
\newblock
\urldef\tempurl%
\url{https://doi.org/10.1145/3183519.3183540}
\showDOI{\tempurl}


\bibitem[\protect\citeauthoryear{Xiong, Wang, Yan, Zhang, Han, Huang, and
  Zhang}{Xiong et~al\mbox{.}}{2017}]%
        {xiong2017precise}
\bibfield{author}{\bibinfo{person}{Yingfei Xiong}, \bibinfo{person}{Jie Wang},
  \bibinfo{person}{Runfa Yan}, \bibinfo{person}{Jiachen Zhang},
  \bibinfo{person}{Shi Han}, \bibinfo{person}{Gang Huang}, {and}
  \bibinfo{person}{Lu Zhang}.} \bibinfo{year}{2017}\natexlab{}.
\newblock \showarticletitle{Precise condition synthesis for program repair}. In
  \bibinfo{booktitle}{\emph{Proceedings of the 39th IEEE/ACM International
  Conference on Software Engineering}}. IEEE, \bibinfo{pages}{416--426}.
\newblock
\urldef\tempurl%
\url{https://doi.org/10.1109/ICSE.2017.45}
\showDOI{\tempurl}


\bibitem[\protect\citeauthoryear{Xuan and Monperrus}{Xuan and
  Monperrus}{2014a}]%
        {xuan2014learning}
\bibfield{author}{\bibinfo{person}{Jifeng Xuan} {and} \bibinfo{person}{Martin
  Monperrus}.} \bibinfo{year}{2014}\natexlab{a}.
\newblock \showarticletitle{Learning to combine multiple ranking metrics for
  fault localization}. In \bibinfo{booktitle}{\emph{Proceedings of the 2014
  IEEE International Conference on Software Maintenance and Evolution}}.
  \bibinfo{pages}{191--200}.
\newblock


\bibitem[\protect\citeauthoryear{Xuan and Monperrus}{Xuan and
  Monperrus}{2014b}]%
        {xuan2014test}
\bibfield{author}{\bibinfo{person}{Jifeng Xuan} {and} \bibinfo{person}{Martin
  Monperrus}.} \bibinfo{year}{2014}\natexlab{b}.
\newblock \showarticletitle{Test case purification for improving fault
  localization}. In \bibinfo{booktitle}{\emph{Proceedings of the 22nd ACM
  SIGSOFT International Symposium on Foundations of Software Engineering}}.
  \bibinfo{pages}{52--63}.
\newblock
\urldef\tempurl%
\url{https://doi.org/10.1145/2635868.2635906}
\showDOI{\tempurl}


\bibitem[\protect\citeauthoryear{Ye, Gu, Martinez, Durieux, and Monperrus}{Ye
  et~al\mbox{.}}{2021}]%
        {ye2019automated}
\bibfield{author}{\bibinfo{person}{He Ye}, \bibinfo{person}{Jian Gu},
  \bibinfo{person}{Matias Martinez}, \bibinfo{person}{Thomas Durieux}, {and}
  \bibinfo{person}{Martin Monperrus}.} \bibinfo{year}{2021}\natexlab{}.
\newblock \showarticletitle{Automated classification of overfitting patches
  with statically extracted code features}.
\newblock \bibinfo{journal}{\emph{IEEE Transactions on Software Engineering}}
  (\bibinfo{year}{2021}).
\newblock


\bibitem[\protect\citeauthoryear{Yuan and Banzhaf}{Yuan and Banzhaf}{2018}]%
        {yuan2018arja}
\bibfield{author}{\bibinfo{person}{Yuan Yuan} {and} \bibinfo{person}{Wolfgang
  Banzhaf}.} \bibinfo{year}{2018}\natexlab{}.
\newblock \showarticletitle{{ARJA:} Automated Repair of Java Programs via
  Multi-Objective Genetic Programming}.
\newblock \bibinfo{journal}{\emph{IEEE Transactions on Software Engineering}}
  (\bibinfo{year}{2018}).
\newblock
\urldef\tempurl%
\url{https://doi.org/10.1109/TSE.2018.2874648}
\showDOI{\tempurl}


\bibitem[\protect\citeauthoryear{Zeller and Hildebrandt}{Zeller and
  Hildebrandt}{2002}]%
        {zeller2002simplifying}
\bibfield{author}{\bibinfo{person}{Andreas Zeller} {and} \bibinfo{person}{Ralf
  Hildebrandt}.} \bibinfo{year}{2002}\natexlab{}.
\newblock \showarticletitle{Simplifying and isolating failure-inducing input}.
\newblock \bibinfo{journal}{\emph{IEEE Transactions on Software Engineering}}
  \bibinfo{volume}{28}, \bibinfo{number}{2} (\bibinfo{year}{2002}),
  \bibinfo{pages}{183--200}.
\newblock


\bibitem[\protect\citeauthoryear{Zhang, Li, Zhang, and Khurshid}{Zhang
  et~al\mbox{.}}{2017b}]%
        {zhang2017boosting}
\bibfield{author}{\bibinfo{person}{Mengshi Zhang}, \bibinfo{person}{Xia Li},
  \bibinfo{person}{Lingming Zhang}, {and} \bibinfo{person}{Sarfraz Khurshid}.}
  \bibinfo{year}{2017}\natexlab{b}.
\newblock \showarticletitle{Boosting spectrum-based fault localization using
  PageRank}. In \bibinfo{booktitle}{\emph{Proceedings of the 26th ACM SIGSOFT
  International Symposium on Software Testing and Analysis}}.
  \bibinfo{pages}{261--272}.
\newblock


\bibitem[\protect\citeauthoryear{Zhang, Gupta, and Gupta}{Zhang
  et~al\mbox{.}}{2006}]%
        {zhang2006locating}
\bibfield{author}{\bibinfo{person}{Xiangyu Zhang}, \bibinfo{person}{Neelam
  Gupta}, {and} \bibinfo{person}{Rajiv Gupta}.}
  \bibinfo{year}{2006}\natexlab{}.
\newblock \showarticletitle{Locating faults through automated predicate
  switching}. In \bibinfo{booktitle}{\emph{Proceedings of the 28th
  International Conference on Software Engineering}}.
  \bibinfo{pages}{272--281}.
\newblock


\bibitem[\protect\citeauthoryear{Zhang, Gupta, and Gupta}{Zhang
  et~al\mbox{.}}{2007}]%
        {zhang2007study}
\bibfield{author}{\bibinfo{person}{Xiangyu Zhang}, \bibinfo{person}{Neelam
  Gupta}, {and} \bibinfo{person}{Rajiv Gupta}.}
  \bibinfo{year}{2007}\natexlab{}.
\newblock \showarticletitle{A study of effectiveness of dynamic slicing in
  locating real faults}.
\newblock \bibinfo{journal}{\emph{Empirical Software Engineering}}
  \bibinfo{volume}{12}, \bibinfo{number}{2} (\bibinfo{year}{2007}),
  \bibinfo{pages}{143--160}.
\newblock


\bibitem[\protect\citeauthoryear{Zhang, Lei, Mao, and Li}{Zhang
  et~al\mbox{.}}{2019}]%
        {zhang2019cnn}
\bibfield{author}{\bibinfo{person}{Zhuo Zhang}, \bibinfo{person}{Yan Lei},
  \bibinfo{person}{Xiaoguang Mao}, {and} \bibinfo{person}{Panpan Li}.}
  \bibinfo{year}{2019}\natexlab{}.
\newblock \showarticletitle{{CNN-FL}: An effective approach for localizing
  faults using convolutional neural networks}. In
  \bibinfo{booktitle}{\emph{Proceedings of the IEEE 26th International
  Conference on Software Analysis, Evolution and Reengineering}}. IEEE,
  \bibinfo{pages}{445--455}.
\newblock


\bibitem[\protect\citeauthoryear{Zhang, Lei, Tan, Mao, Zeng, and Chang}{Zhang
  et~al\mbox{.}}{2017a}]%
        {zhang2017deep}
\bibfield{author}{\bibinfo{person}{Zhuo Zhang}, \bibinfo{person}{Yan Lei},
  \bibinfo{person}{Qingping Tan}, \bibinfo{person}{Xiaoguang Mao},
  \bibinfo{person}{Ping Zeng}, {and} \bibinfo{person}{Xi Chang}.}
  \bibinfo{year}{2017}\natexlab{a}.
\newblock \showarticletitle{Deep learning-based fault localization with
  contextual information}.
\newblock \bibinfo{journal}{\emph{IEICE Transactions on Information and
  Systems}} \bibinfo{volume}{100}, \bibinfo{number}{12} (\bibinfo{year}{2017}),
  \bibinfo{pages}{3027--3031}.
\newblock


\bibitem[\protect\citeauthoryear{Zheng, Hu, and Wang}{Zheng
  et~al\mbox{.}}{2016}]%
        {zheng2016fault}
\bibfield{author}{\bibinfo{person}{Wei Zheng}, \bibinfo{person}{Desheng Hu},
  {and} \bibinfo{person}{Jing Wang}.} \bibinfo{year}{2016}\natexlab{}.
\newblock \showarticletitle{Fault localization analysis based on deep neural
  network}.
\newblock \bibinfo{journal}{\emph{Mathematical Problems in Engineering}}
  \bibinfo{volume}{2016} (\bibinfo{year}{2016}).
\newblock


\bibitem[\protect\citeauthoryear{Zou, Liang, Xiong, Ernst, and Zhang}{Zou
  et~al\mbox{.}}{2019}]%
        {zou2019empirical}
\bibfield{author}{\bibinfo{person}{Daming Zou}, \bibinfo{person}{Jingjing
  Liang}, \bibinfo{person}{Yingfei Xiong}, \bibinfo{person}{Michael~D Ernst},
  {and} \bibinfo{person}{Lu Zhang}.} \bibinfo{year}{2019}\natexlab{}.
\newblock \showarticletitle{An empirical study of fault localization families
  and their combinations}.
\newblock \bibinfo{journal}{\emph{IEEE Transactions on Software Engineering}}
  (\bibinfo{year}{2019}).
\newblock


\end{thebibliography}
